\documentclass[aps,pra,amsmath,amssymb,floatfix,twocolumn,amsmath,superscriptaddress,twocolumn,nofootinbib,tighten,letterpaper]{revtex4-1}
\usepackage{multirow}
\usepackage{bbold}
\usepackage{subfigure}
\usepackage{color}
\usepackage{mathrsfs}
\usepackage{hyperref}
\usepackage[normalem]{ulem}
\usepackage{bm}

\usepackage{amsfonts, relsize, color}
\usepackage{graphics}
\usepackage{graphicx}
\usepackage{subfigure}
\usepackage{hyperref}
\usepackage{color}
\usepackage{comment}

\begin{document}
\title{\bf Dislocation as a bulk probe of higher-order topological insulators}

\author{Bitan Roy}\email{bitan.roy@lehigh.edu}
\affiliation{Department of Physics, Lehigh University, Bethlehem, Pennsylvania, 18015, USA}

\author{Vladimir Juri\v ci\' c}\email{vladimir.juricic@su.se}
\affiliation{Nordita,
KTH Royal Institute of Technology and Stockholm University,
Hannes Alfv\' ens v\" ag 12, SE-106 91 Stockholm, Sweden}
\affiliation{Departamento de F\'isica, Universidad T\'ecnica Federico Santa Mar\'ia, Casilla 110, Valpara\'iso, Chile}

\date{\today}
\begin{abstract}
Topological materials occupy the central stage in the modern condensed matter physics because of their robust metallic edge or surface states protected by the topological invariant, characterizing the electronic band structure in the bulk. Higher-order topological (HOT) states extend this usual bulk-boundary correspondence, so they host the modes localized at lower-dimensional boundaries, such as corners and hinges. Here we theoretically demonstrate that dislocations, ubiquitous defects in crystalline materials, can probe higher-order topology,
recently realized in various platforms. We uncover that HOT insulators respond to dislocations through symmetry protected finite-energy in-gap electronic modes, localized at the defect core, which originate from an interplay between the orientation of the HOT mass domain wall and the Burgers vector of the dislocation. As such, these modes become gapless only when the Burgers vector points toward lower-dimensional gapless boundaries. Our findings are consequential for the systematic probing of the extended bulk-boundary correspondence in a broad range of HOT crystals, and photonic and phononic or mechanical metamaterials through the bulk topological lattice defects.
\end{abstract}

\maketitle

\section{Introduction}

The nontrivial topological invariant characterizing the bulk electronic band structure gives rise to robust edge or surface modes, manifesting the hallmark of a topological material - the bulk-boundary correspondence~\cite{hasan-kane-rmp2010, qi-zhang-rmp2011}. As such, these boundary modes have been so far almost exclusively used to experimentally detect nontrivial electronic topology, both in gapped~\cite{molenkamp-science2007, hsieh-nature2008, xia-natphys2009, hsieh-science2010, chen-science2010, dziawa-natmat2012} and gapless~\cite{xu-science2015, lv-prx2015} systems. Equally important, but much less explored, is the direct probing of topological states in the bulk without invoking the boundary modes, through their response to topological lattice defects, such as \emph{dislocations}~\cite{ran-natphys2009, teo-prb2010, juricic-prl2012, asahi-prb2012, slager-natphys2013, hughes-prb2014, slager-prb2014, you-prb2016, nag-roy-2020, panigrahi-2021}. Moreover, the topological defect modes are more pristine, being immune to contamination by the interfaces and independent of the surface termination. In fact, in the context of experimental probing of topology in the quantum materials this aspect has started to gain prominence only recently~\cite{hamasaki-apl2017, nayak-sciadv2019}.

\begin{figure*}[t!]
\subfigure[]{\includegraphics[width=0.30\linewidth]{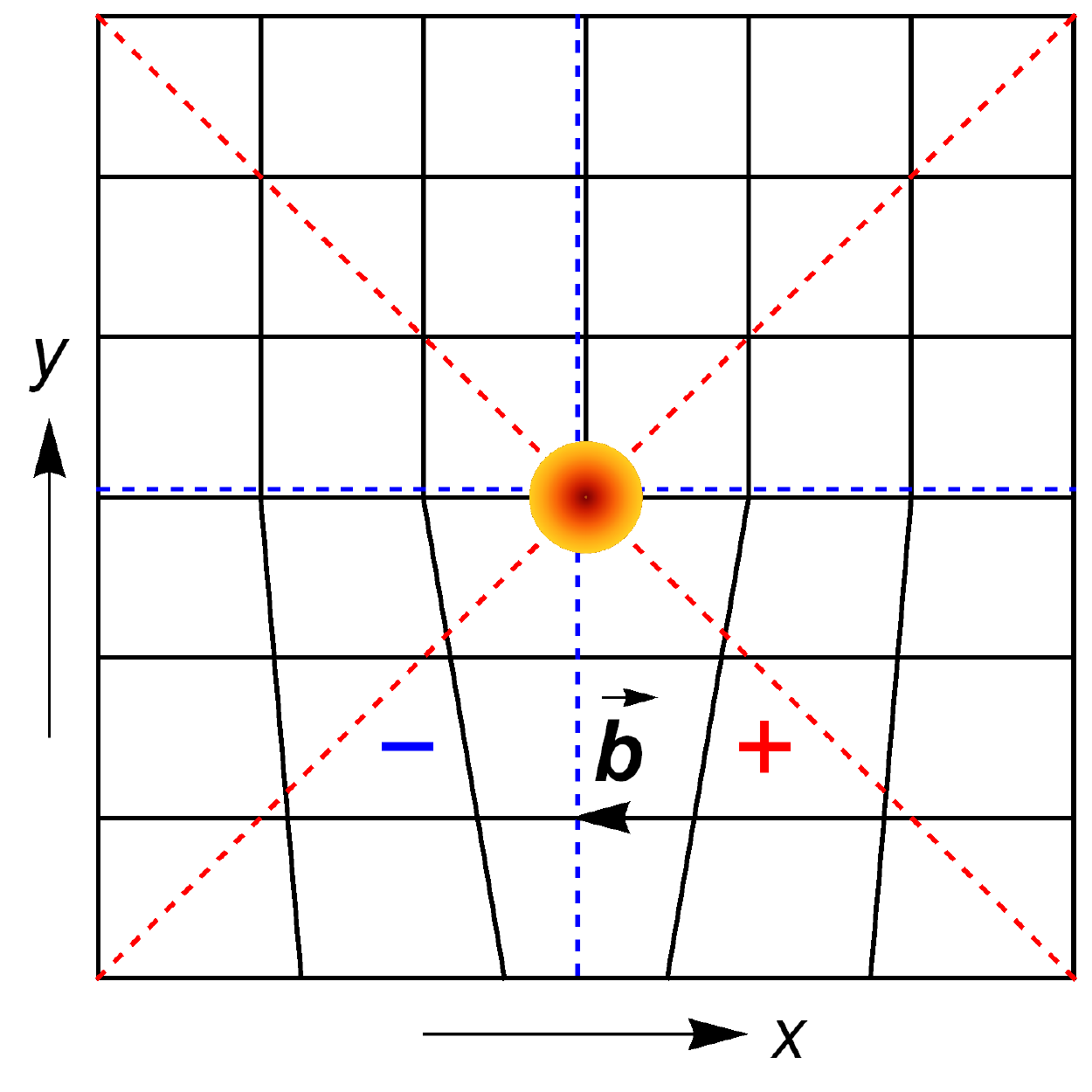}}
\subfigure[]{\includegraphics[width=0.31\linewidth]{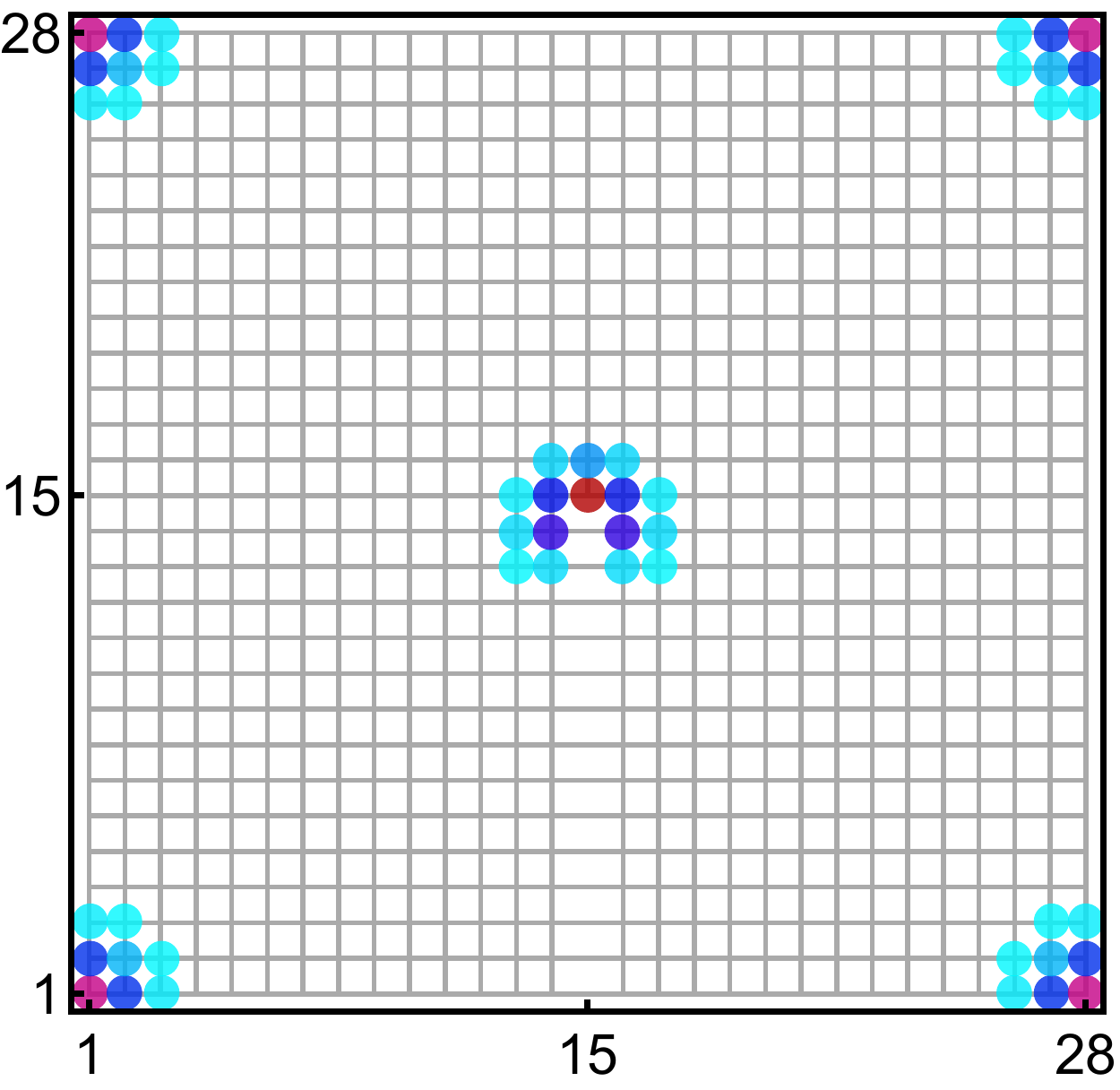}}
\subfigure[]{\includegraphics[width=0.305\linewidth]{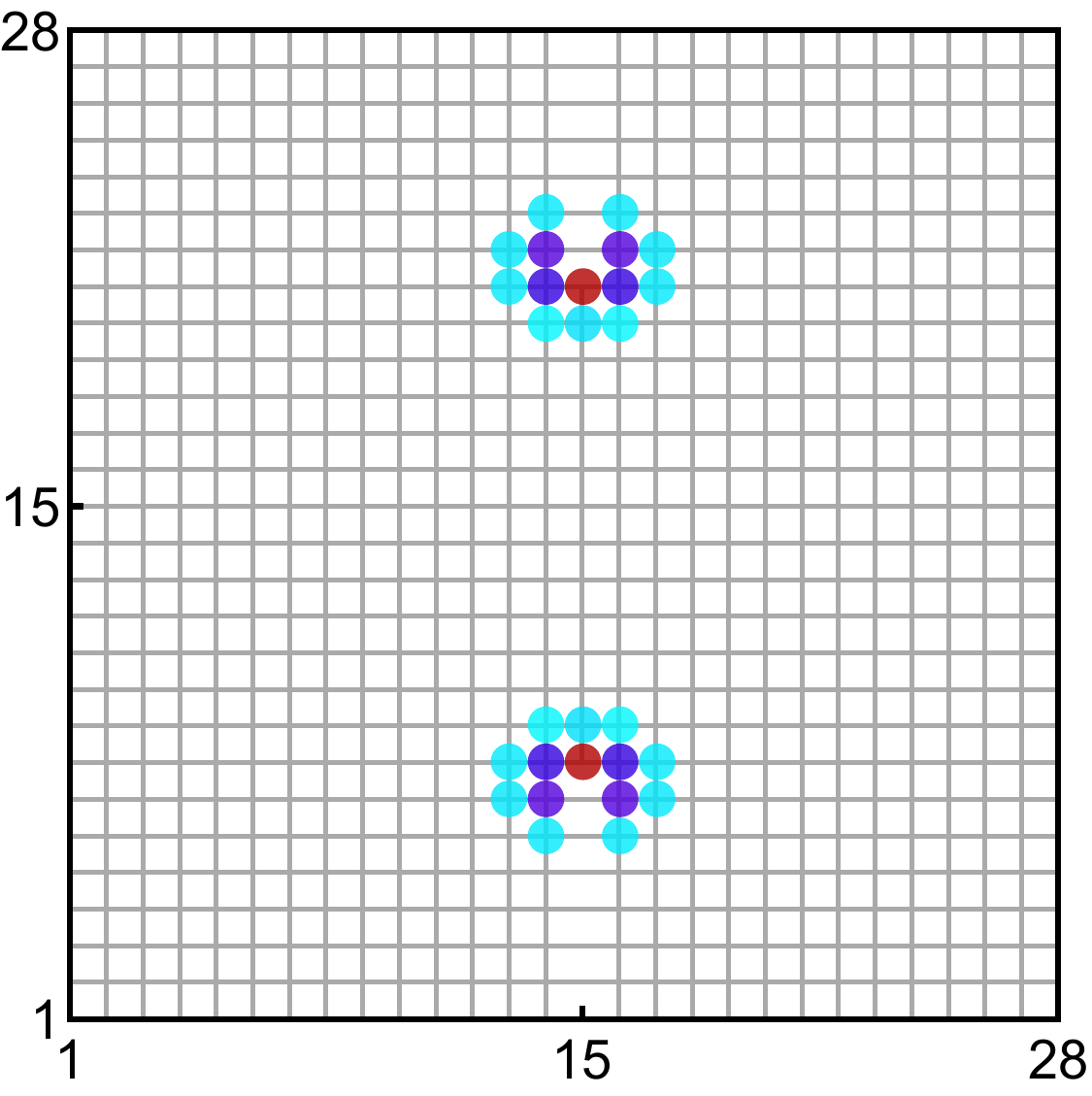}}
\includegraphics[width=0.06\linewidth]{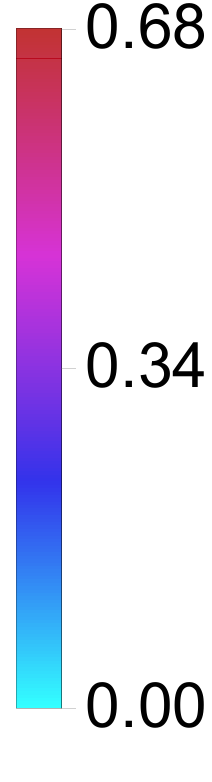}
\caption{Dislocation defect on a square lattice. (a) The defect is obtained through the Volterra cut-and-glue procedure by removing a line of atoms ending at the center of the lattice (orange) and reconnecting the edges across this Volterra cut, which are denoted by $+$ and $-$, right and left from the center, respectively. The corresponding Burgers vector is ${\bf b}=-a{\bf e}_x$. The HOT mass domain walls along (across) which it vanishes (changes sign) are represented by the red and blue dashed lines for $\theta=0$ and $\pi/2$, respectively, as given by Eq.~\eqref{eq:lattice-hamiltonian-HOT}. (b) Local density of states (LDoS) for the dislocation mode localized at the defect core together with the four corner modes in a second-order translationally-active HOT insulator with the band inversion at the $M$ point of the BZ. Here, we set $\theta=0$ in the HOT mass, so that the defect modes are at finite energies, while the corner modes are at zero energy. (c) LDoS for the zero-energy dislocation modes in the periodic system with a dislocation-antidislocation pair for $\theta=\pi/2$. Here, we set $t=2B=1$, $m=3$ and $\Delta=0.20$ [see Eq.~(\ref{eq:firstordermass})]. Any site with LDoS less than $10^{-3}$ is left empty. See also Appendix~\ref{app:D}.}
~\label{Fig:dislocation_defect}
\end{figure*}

\begin{figure*}[t!]
\subfigure[]{\includegraphics[width=0.32\linewidth]{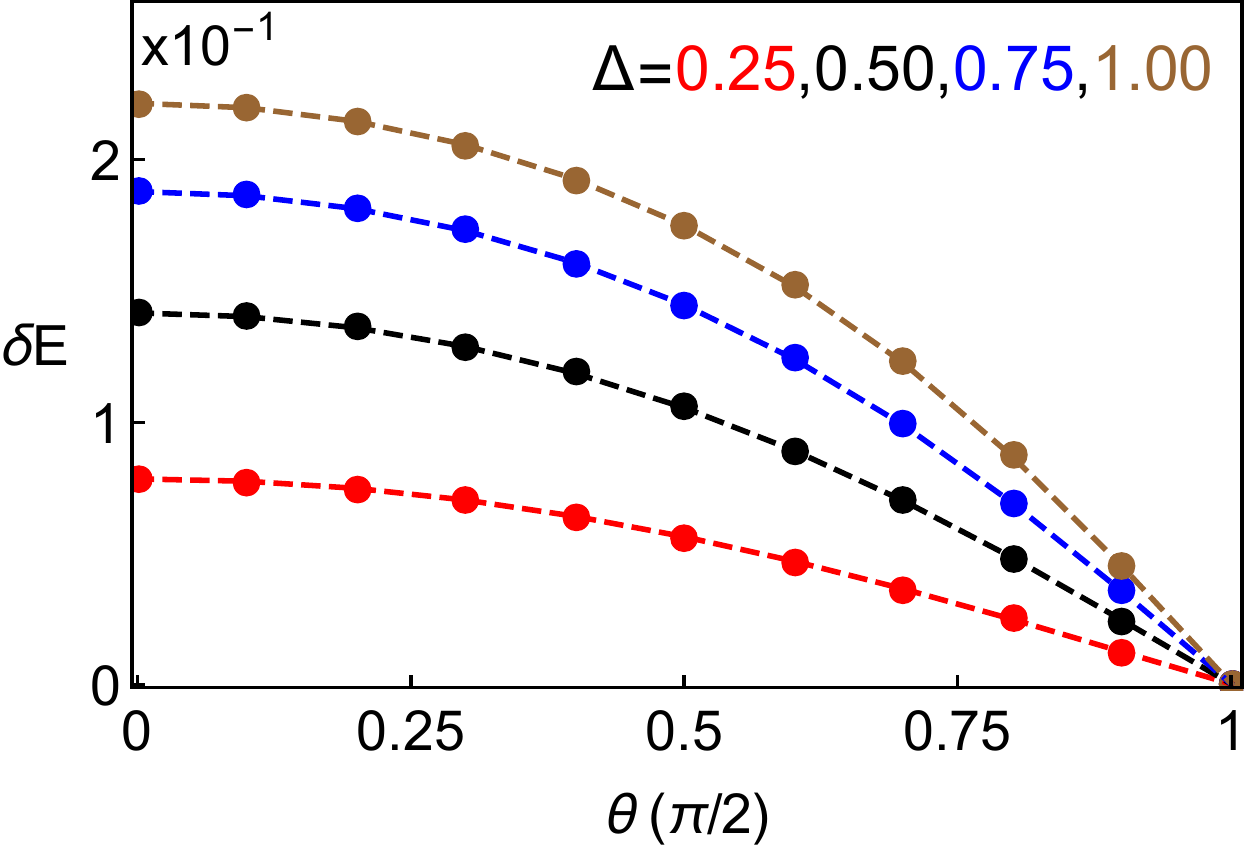}}
\subfigure[]{\includegraphics[width=0.325\linewidth]{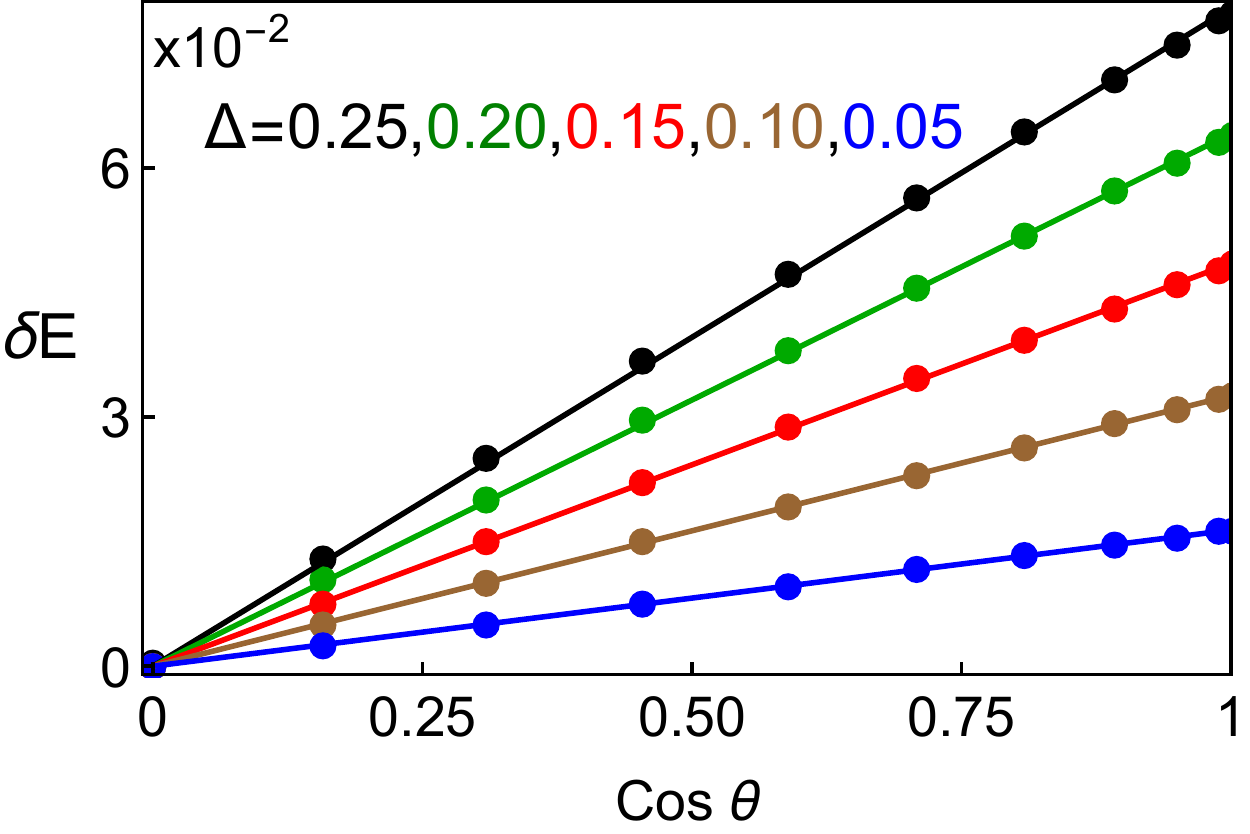}}
\subfigure[]{\includegraphics[width=0.32\linewidth]{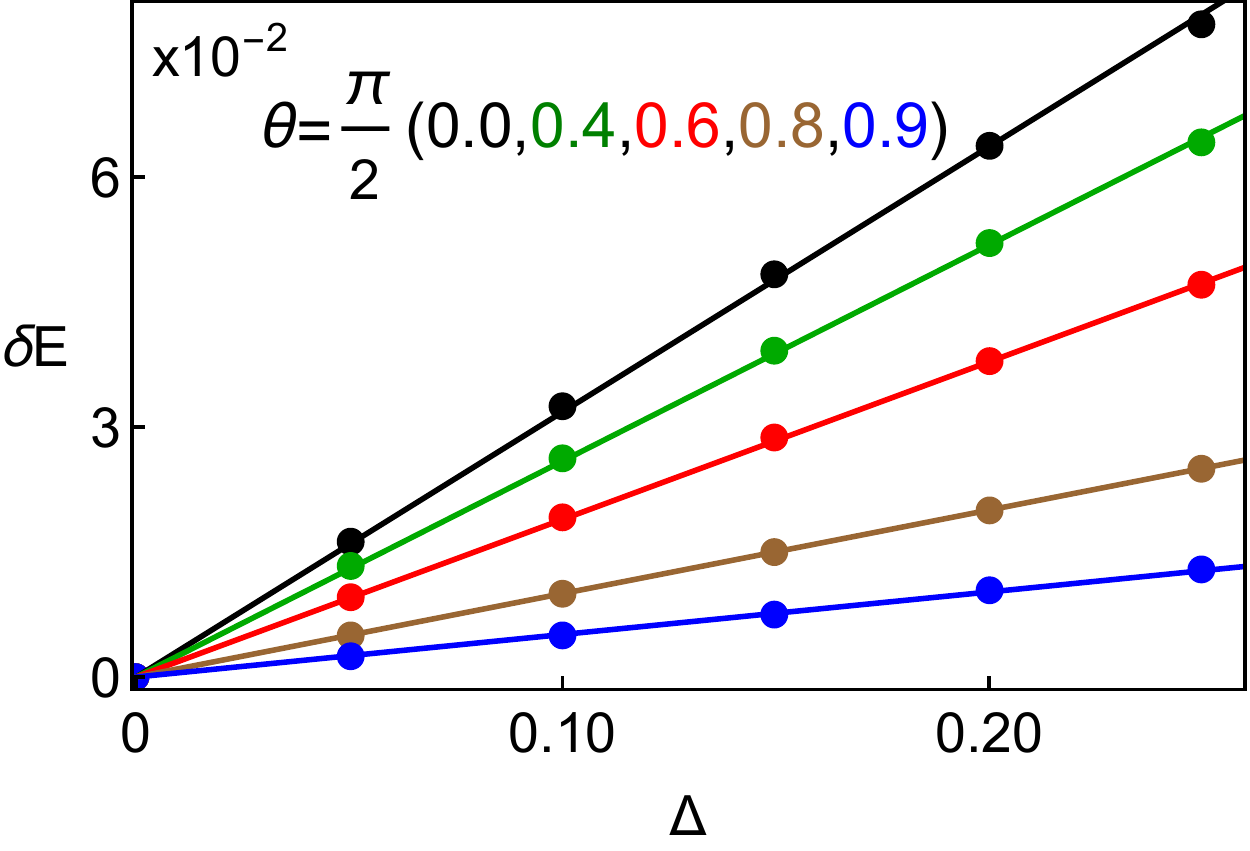}}
\caption{Spectral features of the dislocation modes in a 2D translationally-active HOT insulator for $t=2B=1$ and $m=3$ [see Eqs.~(\ref{eq:lattice-hamiltonian-HOT}) and ~(\ref{eq:firstordermass})]. (a) The scaling of the spectral gap ($\delta E$) among four states localized at the core of the dislocation with $\theta$ measuring the orientation of the four-fold symmetry breaking mass domain wall [see Eq.~\eqref{eq:lattice-hamiltonian-HOT}]. Note that these modes become zero-energy states as $\theta \to \pi/2$, i.e. when the Burgers vector becomes parallel to one of the four-fold symmetry breaking axes [see Fig.~\ref{Fig:dislocation_defect}(a)]. The various choices of the amplitude of the $C_4$ symmetry breaking mass ($\Delta$) are quoted in the figure. For small $\Delta$, the spectral gap ($\delta E$) scales \emph{linearly} with (b) $\cos \theta$ and (c) $\Delta$ [see Eq.~(\ref{eq:Esplitting})]. Consult Appendix~\ref{app:D} for additional details.}
~\label{Fig:dislocation_defect-scaling}
\end{figure*}

In $D-$dimensional $n$th order topological states~\cite{benalcazar-science2017,benalcazar-prb2017,song-prl2017,langbehn-prl2017,schindler-sciadv2018}, bulk probing of the electronic band topology should play an important role, because the extended bulk-boundary correspondence is realized through gapless modes on the lower, $(D-n)-$dimensional boundaries, characterized by codimension $d_c=n$, such as hinges ($d_c=D-1$) and corners ($d_c=D$)~\cite{schindler-natphys2018,serra-nature2018,peterson-nature2018,imhof-natphys2018,mittal-natphot2019, zhang-natphys2019, ni-natmat2019,kempkes-natmat2019}. Their robustness originates from the combination of spatial symmetries, such as discrete rotations, and non-spatial ones, such as the reversal of time. Importantly, these protected modes on lower-dimensional boundaries may be thought of as inherited from the parent, first-order topological state (with $n=1$) upon partially gapping out its edge or surface modes ($d_c=1$), which can, in turn, yield a hierarchical ladder of HOT states~\cite{calugaru-prb2019, nag-juricic-roy-prb2021}. This is accomplished by a suitable \emph{domain wall} mass which changes sign across corners [see Fig.~\ref{Fig:dislocation_defect}(a)] or hinges, thus localizing topological modes at these lower-dimensional boundaries [see Figs.~\ref{Fig:dislocation_defect}(b),~\ref{Fig:dislocation_defect-scaling},~\ref{Fig:3Ddislocation_defect}(a) and ~\ref{Fig:3Ddislocation_defect}(c)]. The reduced dimensionality of the boundary may, however, hinder the experimental detection of the gapless modes, and therefore HOT states require other means to directly probe the bulk electronic topology.

As we demonstrate here, dislocations can serve as bulk probes of HOT insulators through the binding of special topologically protected  electronic modes, see Figs.~\ref{Fig:dislocation_defect} and \ref{Fig:3Ddislocation_defect}. To formulate the mechanism, we recall the Volterra construction:  in a two-dimensional (2D) lattice a dislocation can be created by removing a line ending at the dislocation center (Volterra cut), and reconnecting the sites across this cut so that the translational symmetry is restored away from the defect center (core), see Fig.~\ref{Fig:dislocation_defect}(a). Therefore, any closed loop around the dislocation center features a missing  translation by the Burgers vector ${\bf b}$, which topologically characterizes the defect. As such, a dislocation provides global frustration to the underlying crystalline order, which translates into a nontrivial effect on the electrons hopping on the lattice. Namely, an electron with a momentum ${\bf K}$ when encircling the dislocation picks up a phase equal to ${\rm exp}[{i\Phi_{\rm dis}}]$, with $\Phi_{\rm dis}={\bf K}\cdot {\bf b} \, ({\rm mod}\, 2\pi)$. In particular, for  topological states with the band inversion momentum at ${\bf K}_{\rm inv}$, the hopping phase is $\Phi_{\rm dis}={\bf K}_{\rm inv}\cdot {\bf b} \, ({\rm mod}\, 2\pi)$~\cite{ran-natphys2009}.

%%%%%%%%%%%%%%%%%%%%%%%%%%%%%%%%%%%%%%%%%%%%%%%
%%%%%%%%%%%%%%%%%%%%%%%%%%%%%%%%%%%%%%%%%%%%%%%
%%%%%%%%%%%%%%%%%%%%%%%%%%%%%%%%%%%%%%%%%%%%%%%
%%%%%%%%%%%%%%%%%%%%%%%%%%%%%%%%%%%%%%%%%%%%%%%
%%%%%%%%%%%%%%%%%%%%%%%%%%%%%%%%%%%%%%%%%%%%%%%
%%%%%%%%%%%%%%%%%%%%%%%%%%%%%%%%%%%%%%%%%%%%%%%
%%%%%%%%%%%%%%%%%%%%%%%%%%%%%%%%%%%%%%%%%%%%%%%
%%%%%%%%%%%%%%%%%%%%%%%%%%%%%%%%%%%%%%%%%%%%%%%
%%%%%%%%%%%%%%%%%%%%%%%%%%%%%%%%%%%%%%%%%%%%%%%
%%%%%%%%%%%%%%%%%%%%%%%%%%%%%%%%%%%%%%%%%%%%%%%

To set the stage, recall that a \emph{translationally-active} first-order topological insulator features at least one band inversion at a finite (non-$\Gamma$) momentum in the Brillouin zone (BZ)~\cite{slager-natphys2013} yielding gapless edge states. When $\Phi_{\rm dis}=\pi$ in a translationally-active topological insulator, after encircling a dislocation, the electrons pick up a hopping phase equal to  $\exp({i\pi})=-1$ across the Volterra cut, see Fig.~\ref{Fig:dislocation_defect}(a). In turn, to resolve the frustration in the hopping  introduced by the defect through the nontrivial phase factor, a Kramers pair of zero-energy states gets localized at the dislocation core~\cite{juricic-prl2012}.

%%%%%%%%%%%%%%%%%%%%%%%%%%%%%%%%%%%%%%%%%%%%%%%%%%%%%%%%%%%%%%%%%%%%%%%%%%55
%%%%%%%%%%%%%%%%%%%%%%%%%%%%%%%%%%%%%%%%%%%%%%%%%%%%%%%%%%%%%%%%%%%%%%%%%%%%%%%
%%%%%%%%%%%%%%%%%%%%%%%%%%%%%%%%%%%%%%%%%%%%%%%%%%%%%%%%%%%%%%%%%%%%%%%%%%%%%%%%%%%%%%%%%%%%%%%%%%%%%
%%%%%%%%%%%%%%%%%%%%%%%%%%%%%%%%%%%%%%%%%%%%%%%%%%%%%%%%%%%%%%%%%%%%%%%%%%%%%%%%%%%%%%%%%%%%%

In contrast, a second-order topological insulator features \emph{gapped} edge states that stem from the mass domain wall in the bulk [Figs.~\eqref{Fig:dislocation_defect}(a) and ~\ref{figappend:cornerevolution}]. The domain wall mass gaps out the edge states but only \emph{partially}, in turn producing the topological corner modes through the Jackiw-Rebbi mechanism~\cite{jackiw-rebbi}. Now, when a dislocation is inserted, the defect modes, as we show, still survive [Figs.~\ref{Fig:dislocation_defect}(b),(c)], but, are moved away from zero energy, since the edge states across the Volterra cut are gapped [Fig.~\ref{Fig:dislocation_defect-scaling}(a)]. Importantly, when the orientation of the Burgers vector (${\bf b}$) is parallel to the direction of the domain wall of the HOT mass, the defect modes are pinned at zero energy. Hierarchy of the HOT states in this way directly translates into the \emph{spectral flow} of the dislocation modes, detectable in the tunneling spectroscopy measurements, for instance. The same mechanism is analogously operative for three-dimensional (3D) second-order translationally-active insulators: an edge dislocation hosts gapped modes which become gapless only when the Burgers vector is parallel to the HOT mass domain wall. Furthermore, a screw dislocation hosts gapless propagating modes only when it is orthogonal to a gapless surface or equivalently parallel to the corresponding surface normal [Fig.~\ref{Fig:3Ddislocation_defect} and Fig.~\ref{FigSM:3Ddislocation}]. Finally, we emphasize that the \emph{composite} $C_4 {\mathcal T}$, ${\mathcal P} {\mathcal T}$ and $C_4 {\mathcal P}$ symmetries protecting the zero-energy corner (hinge) modes, also protect both finite- and zero-energy dislocation modes, displayed in Figs.~\ref{Fig:dislocation_defect}(b), ~\ref{Fig:3Ddislocation_defect}(a), and~\ref{Fig:3Ddislocation_defect}(c). Here $C_4$, ${\mathcal T}$ and ${\mathcal P}$ represent discrete four-fold rotational, time-reversal and parity symmetries, respectively. For details consult Appendix~\ref{app:symmetry-protection}, where it is also shown that this protection mechanism for dislocation modes extends to $C_{4n}$ rotational symmetry breaking HOT insulators in two and three dimensions, with $n>1$. Furthermore, in Appendix~\ref{app:F} we show that the defect modes are protected also in the case of $C_{4n+2}$ rotational symmetry breaking HOT insulators, where $n\geq1$.

The rest of the paper is organized as follows. In Sec.~\ref{sec:tightbinding}, we discuss the universal tight-binding model for the HOT insulators in $d=2$ and $d=3$. Section~\ref{sec:2Dnumerics} is devoted to the numerical results for the dislocation modes on a square lattice. In Sec.~\ref{sec:2Danalytic} we present a general argument for the existence of the dislocation modes in a HOT insulator, and in Sec.~\ref{sec:3Ddislocation} we show the numerical results for the dislocation modes in 3D HOT insulators. We discuss the results in Sec.~\ref{sec:discussions} and highlight their possible realizations in HOT crystals and metamaterials. Additional technical details are relegated to the Appendices.

%%%%%%%%%%%%%%%%%%%%%%%%%%%%%%%%%%%%%%%%%%%%%%%%%%%%%%%%%%%%%%%%%%%%%%%%%%%%%%%%%%%%%
%%%%%%%%%%%%%%%%%%%%%%%%%%%%%%%%%%%%%%%%%%%%%%%%%%%%%%%%%%%%%%%%%%%%%%%%%%%%%%%%%%%%%%%

\section{Tight-binding model}~\label{sec:tightbinding}

To show the outlined general mechanism, we take the minimal,  but the universal tight-binding model describing a second-order topological insulator in $d=2$ and $d=3$~\cite{trifunovic-prx2019, calugaru-prb2019}, with the Hamiltonian $H=\sum_{{\bf k}}\Psi_{\bf k}^\dagger \hat{h} \Psi_{\bf k}$, where $\hat{h} = \hat{h}_{0}+\hat{h}_{\Delta}$, and
\allowdisplaybreaks[4]
\begin{align}
\hat{h}_{0}&= {\bf d}({\bf k})\cdot{\bm \Gamma}+M({\bf k})\Gamma_{d+1}, \nonumber\\
\hat{h}_{\Delta}&=\Delta \; \Gamma_{d+2} \; \big\{\cos \theta \left[ \cos(k_x a) - \cos(k_y a) \right]\nonumber\\
&+ \sin \theta \left[ \sin(k_x a)\sin(k_y a) \right] \big\} \equiv\Delta({\bf k},\theta) \; \Gamma_{d+2}.
\label{eq:lattice-hamiltonian-HOT}
\end{align}
Here $({\bm \Gamma},\Gamma_{d+1},\Gamma_{d+2})$ are the mutually anticommuting four-component $\Gamma$ matrices, ${\bf k}$ is the momentum, and $a$ is the lattice spacing. The above Hamiltonian breaks C$_4$ rotational symmetry about the $z-$axis, generated by $R_4=\exp(i\pi\Gamma_{12}/4)$, where $\Gamma_{12}=[\Gamma_1,\Gamma_2]/(2i)$, the time-reversal and the parity symmetries with (representation-dependent) operators ${\mathcal T}$ and ${\mathcal P}$, respectively, but preserves their products $C_4{\mathcal T}$, ${\mathcal T} {\mathcal P}$ and $C_4 {\mathcal P}$. Under four-fold rotation $k_x \to -k_y$ and $k_y \to k_x$. It should be noted that with the above form of the generator of rotations, following the Lie group, the HOT mass \emph{always} breaks discrete rotational symmetry.

\begin{figure*}[t!]
\subfigure[]{\includegraphics[width=0.22\linewidth]{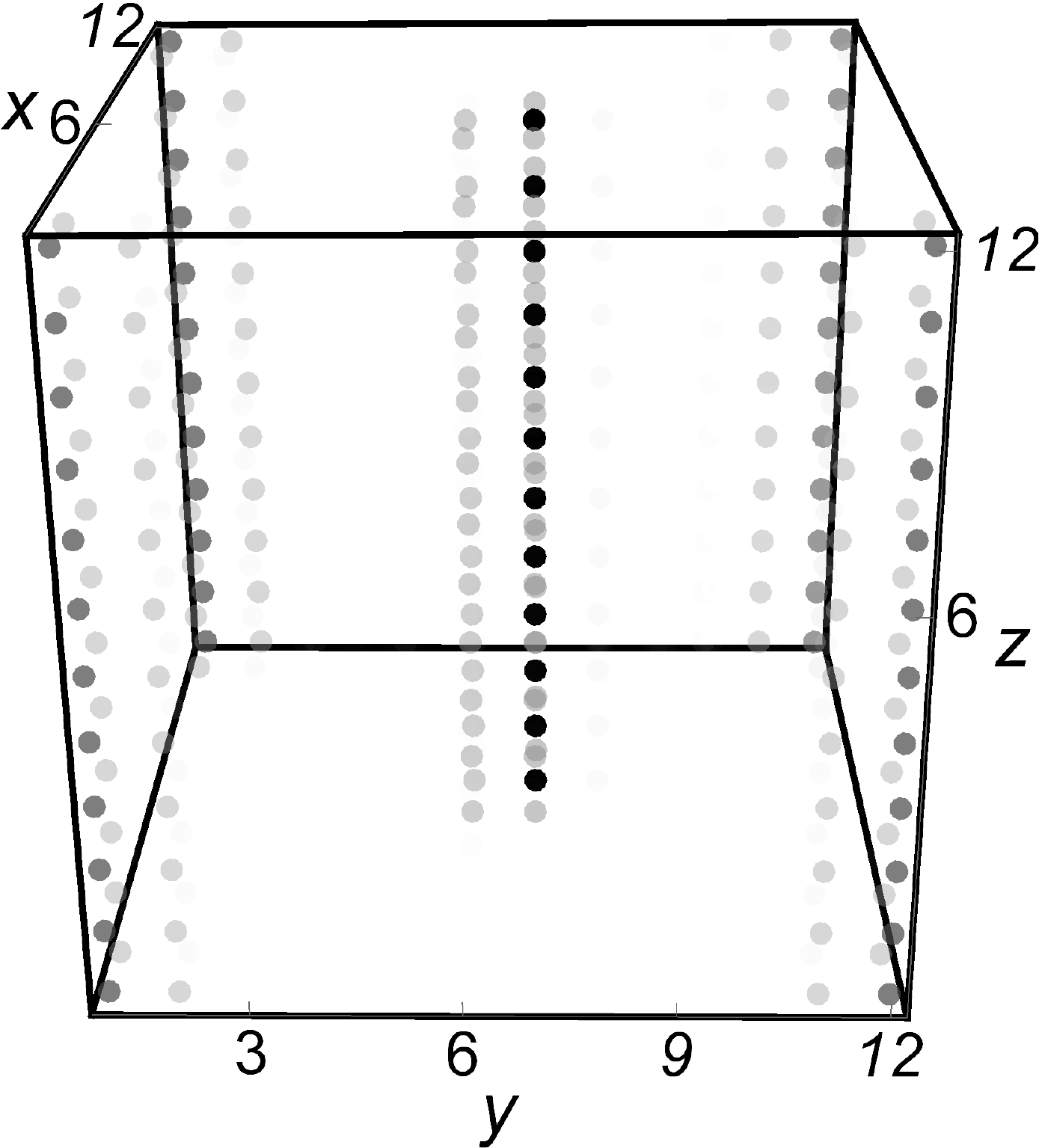}}
\subfigure[]{\includegraphics[width=0.22\linewidth]{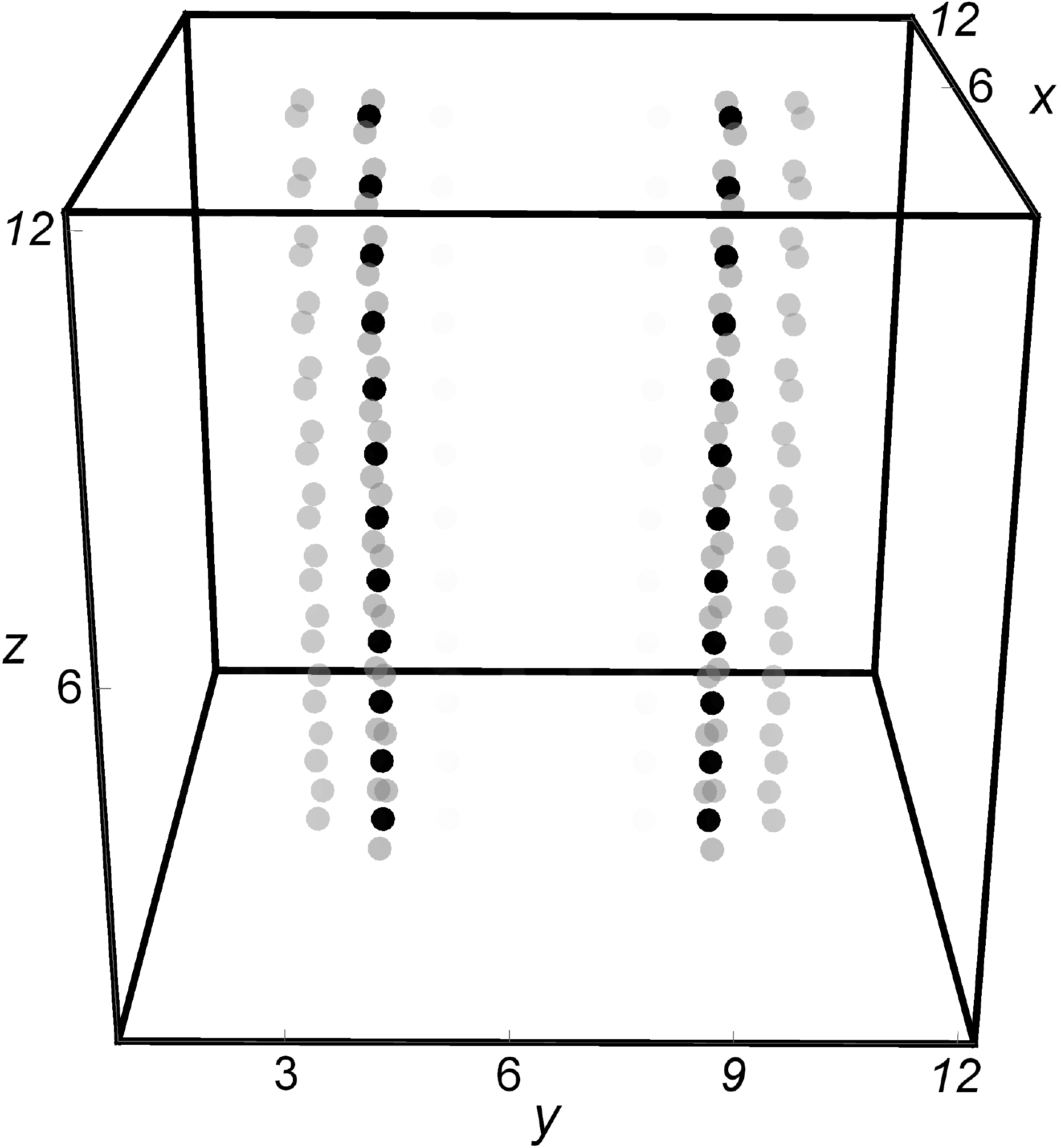}%
\includegraphics[width=0.05\linewidth]{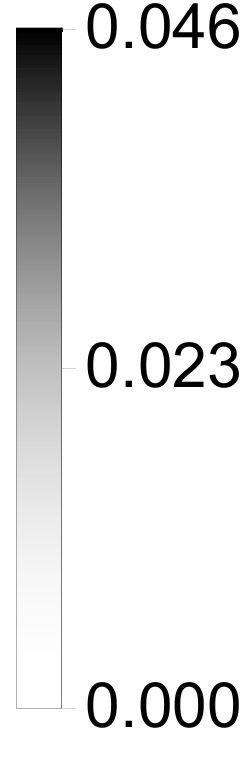}}
\subfigure[]{\includegraphics[width=0.22\linewidth]{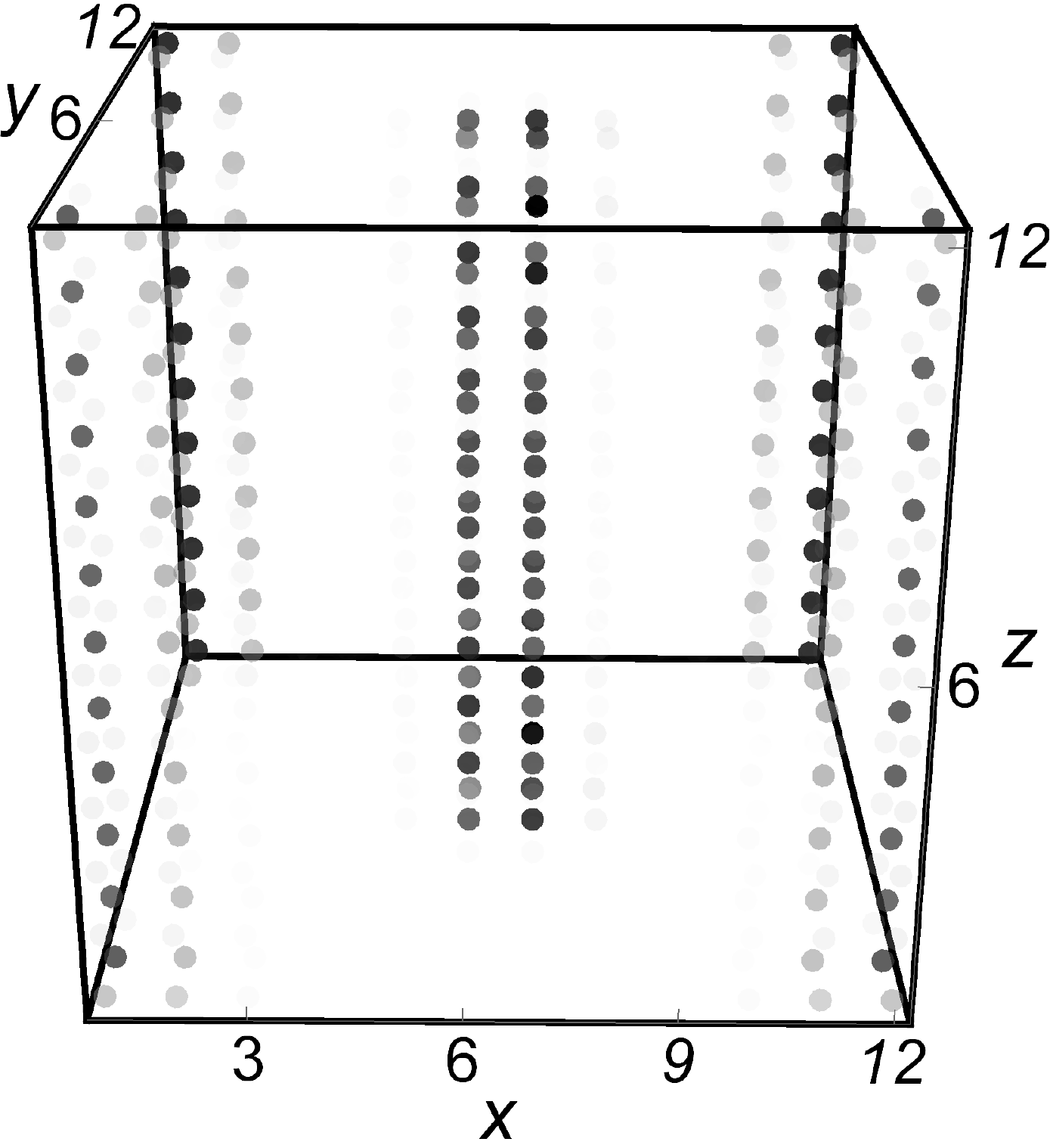}}
\subfigure[]{\includegraphics[width=0.22\linewidth]{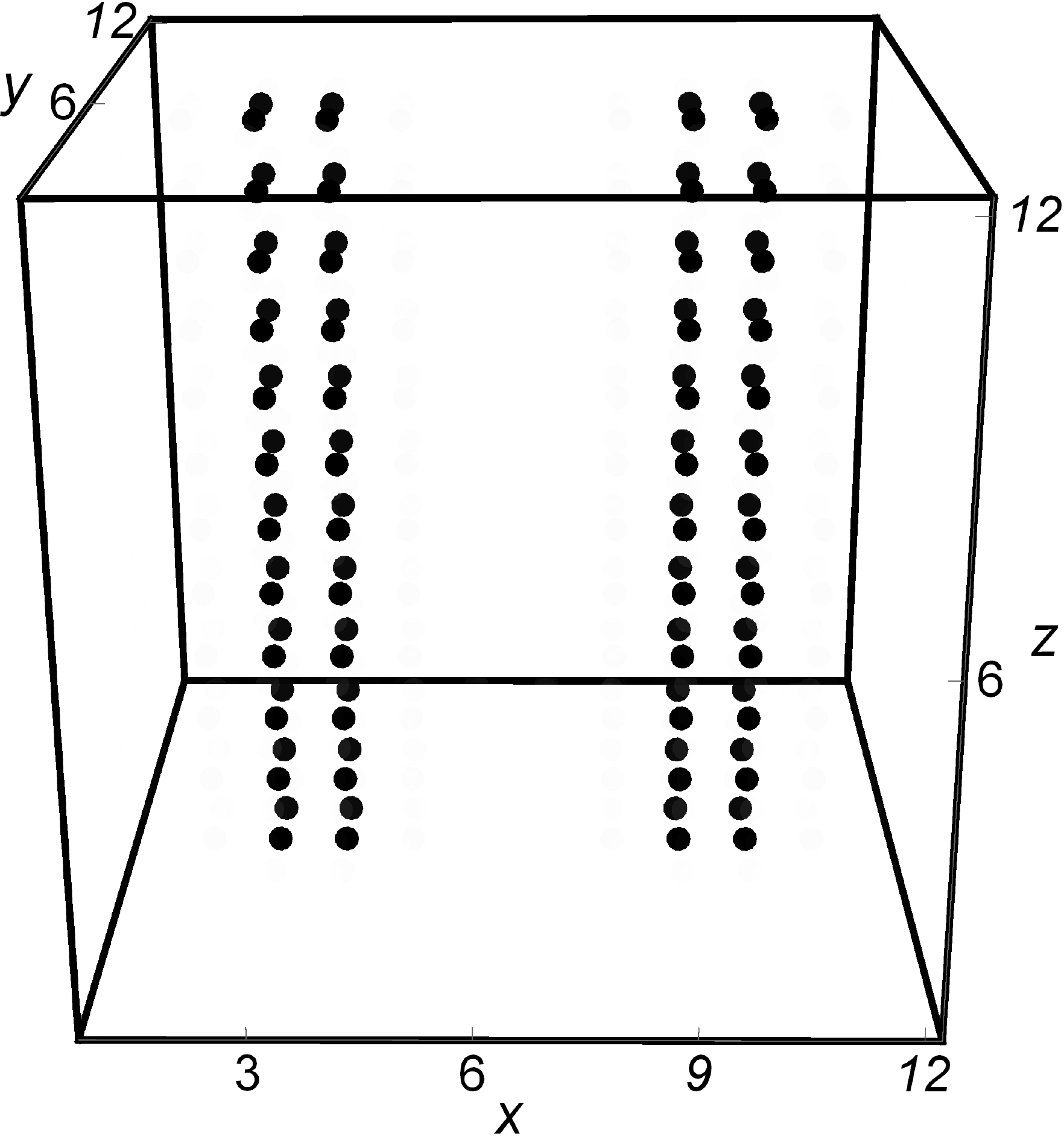}%
\includegraphics[width=0.05\linewidth]{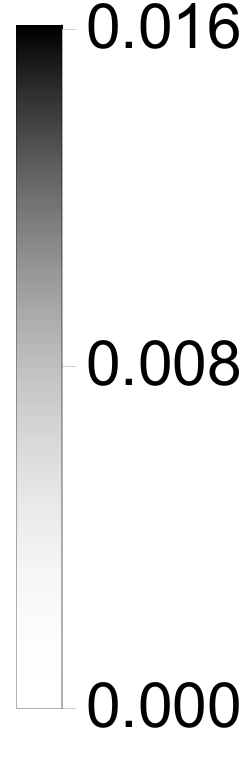}}
\caption{Dislocation defect modes in a 3D HOT insulator on a cubic lattice. LDoS for the \emph{closest} to zero energy modes localized at
(a) an edge dislocation directed in the $z$-direction with the Burgers vector ${\bf b}=a {\bf e}_x$ together with the zero-energy hinge modes,
(b) an edge dislocation-antidislocation pair directed in the $z$-direction  with the Burgers vectors ${\bf b}=\pm a {\bf e}_x$,
(c) a screw dislocation with the Burgers vector ${\bf b}=a {\bf e}_z$ together with the zero-energy hinge modes, and
(d) a screw dislocation-antidislocation pair with the Burgers vectors ${\bf b}=\pm a {\bf e}_z$. Here (a) and (c) [(b) and (d)] are open [periodic] systems. The edge dislocation modes are gapped [see Figs.~\ref{FigSM:3Ddislocation}(a),(b),(c)], while screw dislocation modes are gapless within the numerical accuracy [${\mathcal{O}(10^{-6})}$], as explicitly shown in Fig.~\ref{FigSM:3Ddislocation}(f). In all the cases the system is in a second-order topological $R$ phase with the band inversion at the $R=(1,1,1)\pi/a$ point of the BZ.  The orientation of the HOT mass domain wall is fixed to be along the diagonals $k_y=\pm k_x$, i.e. $\theta=0$ in Eq.~\eqref{eq:lattice-hamiltonian-HOT}. Throughout we set $t=B=1$, $m=10$, and $\Delta=0.40$ [see Eqs.~(\ref{eq:lattice-hamiltonian-HOT}) and (\ref{eq:firstordermass})]. Details of the numerical analysis are presented in Appendix~\ref{app:D}.}
~\label{Fig:3Ddislocation_defect}
\end{figure*}

In both $d=2$ and $d=3$, we take the form factors $d_i({\bf k})=t \sin k_i a$, while the first-order mass is given by
\begin{align}~\label{eq:firstordermass}
M({\bf k}) &= m - 2B \bigg[ d- \sum^{d}_{j=1}\cos(k_j a) \bigg].
\end{align}
We consider translationally-active $M$ phase in $d=2$ with a band inversion at $M=(\pi/a,\pi/a)$ point in the BZ, which is realized in the parameter range $4 < m/B < 8$. In $d=3$ we take translationally-active $R$ phase with the band inversion at $R$ point in the BZ, $R=(\pi/a,\pi/a,\pi/a)$, for $8 < m/B < 12$. The resulting first-order topological insulator supports edge and surface states, both with $d_c=1$,  respectively, in $d=2$ and $d=3$. Furthermore, $\hat{h}_{\Delta}$ acts as a mass term for the topological edge (surface) states, and leaves only the corners (hinges) gapless, yielding a second-order topological insulator with corner (hinge) modes with $d_c=2$ [Figs.~\ref{Fig:dislocation_defect}(b),~\ref{Fig:3Ddislocation_defect}(a),~\ref{Fig:3Ddislocation_defect}(c)]. To be concrete, in Eq.~\eqref{eq:lattice-hamiltonian-HOT} we fix this Wilson-Dirac mass term so that it changes sign under the $C_4$ rotation, transforming $k_x\to -k_y$, $k_y\to k_x$. As such, this mass term for each value of the parameter $0\leq\theta\leq\pi/2$ necessarily features a line across which it changes sign. In particular, for $\theta=\pi/2$ the domain wall lies along the principal axes, $k_x=0, k_y=0$, while for $\theta=0$ it is located along the diagonals $k_y=\pm k_x$. See Fig.~\ref{Fig:dislocation_defect}(a) for the HOT mass domain walls in the real space.

%%%%%%%%%%%%%%%%%%%%%%%%%%%%%%%%%%%%%%%%%%%%%%%%%%%%%%%%%%%%%%%%%%%%%%%%%%%%%%%%%%%%%%%%%%%%%%%
%%%%%%%%%%%%%%%%%%%%%%%%%%%%%%%%%%%%%%%%%%%%%%%%%%%%%%%%%%%%%%%%%%%%%%%%%%%%%%%%%%%%%%%%%%%%%%%%%

\section{2D Lattice dislocation modes}~\label{sec:2Dnumerics}

We perform numerical analysis of the translationally-active HOT $M$ phase, hosting a band inversion at the $M=(\pi/a,\pi/a)$ point in the BZ. The implementation of the model, given by Eq.~\eqref{eq:lattice-hamiltonian-HOT}, was carried out in the real space on a square lattice hosting a dislocation defect with the Burgers vector ${\bf b}=a {\bf e}_x$, oriented in the lattice $x-$direction, as shown in Fig.~\ref{Fig:dislocation_defect}(a). See also Appendix~\ref{app:D} for details. In an open system, the dislocation modes and the corner states coexist in the HOT $M$ phase, explicitly showing that the defect can probe the extended bulk-boundary correspondence, see Fig.~\ref{Fig:dislocation_defect}(b). Furthermore, we find that dislocations bind the modes in a lattice without boundaries (periodic system), further corroborating their role as a pure bulk probe of higher-order electronic topology, see Fig.~\ref{Fig:dislocation_defect}(c). The hybridization effects in both cases can be neglected, as the defect modes are localized within a few lattice sites around its center, which is much shorter than both the system size and the separation between the defects.

%%%%%%%%%%%%%%%%%%%%%%%%%%%%%%%%%%%%%%%%%%%%%%
%%%%%%%%%%%%%%%%%%%%%%%%%%%%%%%%%%%%%%%%%%%%%%
%%%%%%%%%%%%%%%%%%%%%%%%%%%%%%%%%%%%%%%%%%%%%%
%%%%%%%%%%%%%%%%%%%%%%%%%%%%%%%%%%%%%%%%%%%%%%
%%%%%%%%%%%%%%%%%%%%%%%%%%%%%%%%%%%%%%%%%%%%%%
%%%%%%%%%%%%%%%%%%%%%%%%%%%%%%%%%%%%%%%%%%%%%%
%%%%%%%%%%%%%%%%%%%%%%%%%%%%%%%%%%%%%%%%%%%%%%
%%%%%%%%%%%%%%%%%%%%%%%%%%%%%%%%%%%%%%%%%%%%%%
%%%%%%%%%%%%%%%%%%%%%%%%%%%%%%%%%%%%%%%%%%%%%%
%%%%%%%%%%%%%%%%%%%%%%%%%%%%%%%%%%%%%%%%%%%%%%

Most importantly, for any choice of the domain wall orientation ($\theta$), we find that the defects feature mid-gap bound states at \emph{finite} energies, see Fig.~\ref{Fig:dislocation_defect-scaling}(a). As the domain wall orientation approaches the direction of the Burgers vector ($\theta\to\pi/2$), the spectral gap ($\delta E$) between the dislocation modes decreases. Eventually, when the two directions coincide ($\theta=\pi/2$), the modes become degenerate zero energy states. For the spectral flow of the dislocation modes, see Fig.~\ref{FigSM:soectralflow2D}. Also notice that finite energy dislocation modes are particle-hole partners, while they become eigenmodes of the particle-hole operator when pinned at zero energy. See Appendix~\ref{secSI:particlehole} for details. The scaling of the gap with the domain wall orientation for various amplitudes of the Wilson-Dirac mass ($\Delta$) is displayed in Fig.~\ref{Fig:dislocation_defect-scaling}(a), showing that $\delta E \to 0$ as $\theta \to \pi/2$. We next present a general argument supporting this observation.

%%%%%%%%%%%%%%%%%%%%%%%%%%%%%%%%%%%%%%%%%%%%%%%%%%%%%%%%%%%%%%%%%%%%%%%%%%%%%%%%%%%%%%%%%%%%%%%%%%%%%%%%%%%%
%%%%%%%%%%%%%%%%%%%%%%%%%%%%%%%%%%%%%%%%%%%%%%%%%%%%%%%%%%%%%%%%%%%%%%%%%%%%%%%%%%%%%%%%%%%%%%%%%%%%%%%%%%%%

 \section{Dislocation modes: a general argument}~\label{sec:2Danalytic}

 To this end, we recall  that in the parent first-order topological insulator before reconnecting the edges across the Volterra cut, each edge (1) features a Kramers pair of zero energy (due to a unitary particle-hole symmetry, see Appendix~\ref{secSI:particlehole}) helical modes, and (2) is perpendicular to the Burgers vector ${\bf b}$. The zero energy states at each of the edges are then the eigenstates of the matrix $A_{\bf b}=i\Gamma_{\bf b}\Gamma_{3}$, where $\Gamma_{\bf b}={\bm \Gamma}\cdot {\bf b}/|{\bf b}|$, and $\Gamma_3$ is the mass matrix for the first-order topological insulator in $d=2$ [see Eq.~(\ref{eq:lattice-hamiltonian-HOT}) and (\ref{eq:firstordermass})], as explicitly shown in the Appendix~\ref{app:A}. The dislocation defect, which is created by the Volterra construction, with the associated hopping $\pi$ phase factor, gives rise to the level repulsion among four zero modes at the \emph{pasted} edges across the Volterra cut. However, a Kramers pair of modes $|\Psi_0\rangle$  still remains pinned at zero energy and gets localized in the defect core~\cite{ran-natphys2009}.

The crucial observation is  that the HOT mass matrix $\Gamma_4$ \emph{commutes} with the dislocation or edge-mode matrix $A_{\bf b}$, $[\Gamma_4,A_{\bf b}]=0$. The HOT mass matrix therefore reduces in the eigen-subspaces of $A_{\bf b}$, introducing the level repulsion between the two zero modes and thus symmetrically splits them about the zero energy. Furthermore, this implies that the modes do not change the form, i.e. they remain localized around at the defect, after introducing the HOT mass. The energy splitting  reads as
\begin{equation}
\delta E=\langle\Psi_0| \Delta[{\bf K}_{\rm inv}-i{\bf b}({\bf b}\cdot{\nabla})]|\Psi_0\rangle.
\end{equation}
Unless $\Delta[{\bf K}_{\rm inv}-i{\bf b}({\bf b}\cdot{\nabla})]=0$, i.e. when the HOT mass vanishes in the direction of the Burgers vector, the energy splitting is non-zero ($\delta E\neq 0$). Therefore, the interplay between the orientation of the HOT mass domain wall and the Burgers vector pins the dislocation modes precisely at zero energy when the two directions are \emph{parallel} [see Fig.~\ref{Fig:dislocation_defect}(a)]. The mechanism for the splitting of the dislocation modes is also operative for gapping out the edges, ultimately yielding the corner modes, implying that a dislocation can directly probe higher-order bulk-boundary correspondence.

This mechanism captures the existence of the localized pair of dislocation modes in the $M$ phase,  split by the energy gap
\begin{equation}~\label{eq:Esplitting}
\delta E=2\Delta \cos\theta {\tilde E},
\end{equation}
where
\begin{equation}
{\tilde E}\sim \left|\int dx \left( \Psi_0^{(1,2)}(x) \right)^\dagger \partial_x^2 \Psi_0^{(1,2)}(x) \right|, \nonumber
\end{equation}
and $\Psi_0^{(1,2)}(x)$ are the zero-energy dislocation states in the first-order phase (see Appendix~\ref{app:A}).
The obtained energy gap ($\delta E$) implies that the modes are pinned at zero energy only when the Burgers vector is parallel with the mass domain wall ($\theta=\pi/2$), see Fig.~\ref{Fig:dislocation_defect-scaling}(a). Furthermore, for small $\Delta$, $\delta E$ scales linearly with $\cos \theta$ [see Fig.~\ref{Fig:dislocation_defect-scaling}(b)] and $\Delta$ [see Fig.~\ref{Fig:dislocation_defect-scaling}(c)].
Finally, a dislocation does not feature any bound states either when $\Phi_{\rm dis}=0$ (as in the $\Gamma$ phase) or in the trivial phase.

%%%%%%%%%%%%%%%%%%%%%%%%%%%%%%%%%%%%%%%%%%%%%%%%%%%%%%%%%%%%%%%%%%%%%%%%%%%%%%%%%%%%%%%%%%%%%%%%%%%%%%%%%%%%%%%%%%%%%%%%
%%%%%%%%%%%%%%%%%%%%%%%%%%%%%%%%%%%%%%%%%%%%%%%%%%%%%%%%%%%%%%%%%%%%%%%%%%%%%%%%%%%%%%%%%%%%%%%%%%%%%%%%%%%%%%%%%%%%%%%%

\section{Dislocations in 3D HOT insulator}~\label{sec:3Ddislocation}

The above arguments can be straightforwardly extended to 3D second-order topological insulators. Notice first that a 3D edge dislocation can be obtained from its 2D analogue by translating it along an out-of-plane lattice vector~\cite{chaikin-book}. Therefore, an edge dislocation in a 3D second-order topological insulator should in general feature \emph{gapped} modes, which, however, become \emph{gapless} for the Burgers vector parallel to the HOT mass domain wall direction, analogously to the 2D case [see Fig.~\ref{FigSM:3Ddislocation}(a)].

A screw dislocation,  being a true 3D defect, features the Burgers vector ${\bf b}$ parallel to its orientation. An electron encircling the dislocation defect once, skips a lattice distance $|{\bf b}|$ along the defect relative to the perfect crystal~\cite{chaikin-book}. The ``${\bf K}\cdot{\bf b}$" rule implies that  an electron then picks up a phase $\Phi_{\rm dis}={\bf K}_{\rm inv}\cdot{\bf b}$. When this phase is nontrivial in a translationally-active first-order insulator, the screw dislocation hosts gapless propagating modes~\cite{ran-natphys2009}. On the other hand, in a translationally-active HOT insulator some of the surfaces are gapped, and the screw dislocation hosts gapless propagating modes \emph{only} when it is oriented perpendicular to gapless surfaces. See Appendix~\ref{SMsec:3Ddislocation}. In particular, in a 3D second-order topological insulator, with a single mass domain wall [see Eq.~(\ref{eq:lattice-hamiltonian-HOT})], a screw dislocation parallel to it hosts gapless modes, since the defect then pierces gapless $xy$ surfaces. Otherwise, a dislocation perpendicular to gapped $xz$ or $yz$ surfaces features gapped one-dimensional modes. See Fig.~\ref{FigSM:3Ddislocation}(f).

We numerically confirm this scenario in a 3D second-order topological insulator described by the tight-binding model, exemplifying the $R-$phase on the cubic lattice (see  Eq.~\eqref{eq:lattice-hamiltonian-HOT} and Appendix~\ref{app:D}). Both single edge dislocation with Burgers vectors ${\bf b}=a{\bf e}_x$ extending in the $z-$direction in an open system, which is also the propagation direction of the zero-energy hinge modes, and a edge dislocation-antidislocation pair extending in the same direction with Burgers vectors ${\bf b}=\pm a{\bf e}_x$ in a periodic system indeed yield finite energy states when $\theta \neq \pi/2$, as shown in  Fig.~\ref{Fig:3Ddislocation_defect}(a) and Fig.~\ref{Fig:3Ddislocation_defect}(b), respectively. For $\theta=\pi/2$ the dislocation modes become gapless (same as in 2D). See Figs.~\ref{FigSM:3Ddislocation}(a),(b),(c). On the other hand, for a single screw dislocation with ${\bf b}=a{\bf e}_z$ in a open system (coexisting with the zero-energy hinge modes) and a screw dislocation-antidislocation pair with ${\bf b}=\pm a{\bf e}_z$ in a periodic system, we obtain \emph{gapless} dislocation modes for any HOT mass domain wall orientation $\theta$ in Eq.~\eqref{eq:lattice-hamiltonian-HOT}, as displayed in Figs.~\ref{Fig:3Ddislocation_defect}(c) and~\ref{Fig:3Ddislocation_defect}(d). See also Fig.~\ref{FigSM:3Ddislocation}(f) where vanishing of the gap is explicitly shown. The propagating modes are gapless in this case because the defect pierces the gapless $xy$ surfaces, for any $\theta$. The dislocation modes are localized within a few lattice sites at the defect core, as can be seen from their LDoS in a plane perpendicular to the dislocation direction displayed in Figs.~\ref{FigSM:3Ddislocation}(d) and \ref{FigSM:3Ddislocation}(e). The screw dislocation modes also inherit the $C_4$ symmetry preserved by the  defect.

%%%%%%%%%%%%%%%%%%%%%%%%%%%%%%%%%%%%%%%%%%%%%%%%%%%%%%%%%%%%%%%%%%%%%%%%%%%%%%%%%%%%%%%%%%%%%%%%%%%%%%%%%%%%%%%%%
%%%%%%%%%%%%%%%%%%%%%%%%%%%%%%%%%%%%%%%%%%%%%%%%%%%%%%%%%%%%%%%%%%%%%%%%%%%%%%%%%%%%%%%%%%%%%%%%%%%%%%%%%%%%%%%%%%%%%

\section{Discussion and conclusion}~\label{sec:discussions}

Our findings are experimentally consequential for probing HOT insulators, apart from the crystalline systems,
also in metamaterials. The paradigmatic model of the 2D second-order topological insulator, the Benalcazar-Bernevig-Hughes (BBH) model~\cite{benalcazar-science2017}, equivalent to the minimal lattice model in Eq.~\eqref{eq:lattice-hamiltonian-HOT}~\cite{roy-prr2019}, has been realized in the lattice of microwave resonators~\cite{peterson-nature2018}. A dislocation defect in this setup should be created by a local hopping modification through $\pi$ phase factors across a line of missing sites ending at the dislocation center, analogously to the case of a translationally-active first-order topological insulator~\cite{grinberg-arxiv2019}, and a disclination~\cite{peterson-arxiv2020}. In the BBH photonic lattice, where the sign of the hopping also can be locally manipulated~\cite{mittal-natphot2019}, it should be therefore possible to introduce the dislocation defects and observe the defect modes, as in first-order 2D topological photonic crystals~\cite{noh-natphot2018,li-natcomm2018}, and for a disclination defect~\cite{liu-arxiv2020}. Finally, the artificial lattices can host HOT phases, as recently shown for Kagome lattice~\cite{kempkes-natmat2019}, and we expect that because of their tunability, our theoretical predictions can also be directly tested in these platforms.

Most of the proposed and experimentally studied 3D HOT crystalline materials turn out to be of the translationally-active type. For example, elemental Bi exhibits a double band inversion at the $C_3$-symmetric $T$-point in the BZ  and supports mixed electronic topology manifesting through coexisting gapless hinge and Dirac surface modes~\cite{schindler-natphys2018,nayak-sciadv2019,hsu-pnas2019}. Our general mechanism thus implies that an edge dislocation  with the Burgers vector in the $(111)$ direction parallel to the HOT mass domain wall (see, Fig.~1c in Ref.~\cite{schindler-natphys2018}), so that also ${\Phi_{\rm dis}}=\pi$, features gapless modes, protected by $C_3$ and time-reversal symmetries. A screw dislocation oriented in the same direction should host one-dimensional gapless (gapped) states if the $(111)$ surface is gapless (gapped). Similarly, for a recently proposed HOT insulator in Zr(TiH$_2$)$_2$, with band-inversion away from the $\Gamma$ point and the gapless modes along all the edges in the cubic geometry~\cite{zhang-nature2019}, we predict gapless (gapped) modes in the core of an edge (a screw) dislocation with the Burgers vector along a principal crystal axis. Finally, the candidate HOT insulators Bi$_4$X$_4$, with X=Br,I,~\cite{zhang-nature2019,vergniory-nature2019,tang-nature2019,yoon-arxiv2020}  feature band inversions at $R$ and $M=(\pi/a,\pi/a,0)$ points in the BZ, and hence the dislocations should host the one-dimensional modes, following the above general rule.

Here we demonstrated that dislocations can be instrumental in probing  higher-order electronic topology in insulators as a consequence of the subtle interplay between the geometry of the HOT Wilson-Dirac mass in the momentum space and real-space lattice topological defects. We furthermore demonstrate the protection of the dislocation modes in the case of $C_{2n}$ rotational symmetry breaking HOT insulators in both two and three dimensions (see Appendices~\ref{app:symmetry-protection} and~\ref{app:F}), which pertains to physically relevant $C_4$ and $C_6$ symmetric crystals. On the other hand, when the order of the rotation is \emph{odd}, i.e. for $C_{2n+1}$ rotations, there is no higher-order Wilson-Dirac mass term that changes sign under such rotation. Therefore, we cannot find second-order topological mass. Further analysis of this case is left for future investigation. Recently, it has been shown that also partial dislocations with a Burgers vector which is a fraction of a primitive lattice vector can host gapless propagating modes in 3D HOT insulators~\cite{Queiroz-PRL2019}. In addition, disclination can also host topological modes in 3D HOT insulators~\cite{Geier-2021}. Our findings motivate future investigation of the response to the dislocations in HOT insulators on different crystalline lattices, such as, for instance, Kagome lattice~\cite{ezawa-PRL2018}.  Finally, we expect that our mechanism will be a useful guide for the experimental detection of the HOT phases in diverse platforms, and consequential also for  HOT semimetals~\cite{calugaru-prb2019, hughes-HOTDSM-PRB2018} and superconductors.

%%%%%%%%%%%%%%%%%%%%%%%%%%%%%%%%%%%%%%%%%%%%%%%
%%%%%%%%%%%%%%%%%%%%%%%%%%%%%%%%%%%%%%%%%%%%%%%
%%%%%%%%%%%%%%%%%%%%%%%%%%%%%%%%%%%%%%%%%%%%%%%
%%%%%%%%%%%%%%%%%%%%%%%%%%%%%%%%%%%%%%%%%%%%%%%
%%%%%%%%%%%%%%%%%%%%%%%%%%%%%%%%%%%%%%%%%%%%%%%

\acknowledgements

B.R. was supported by the Startup Grant from Lehigh University. V.J. acknowledges support of the Swedish Research Council (VR 2019-04735).

\begin{center}
{\bf Data and code availability}
\end{center}
The data that support the plots within this paper and other findings of this study are available from the authors upon reasonable request.

%%%%%%%%%%%%%%%%%%%%%%%%%%%%%%%%%%%%%%%%%%%%%%%%%%%%%%%%%%%%%%%%%%%%%%%%%%%%%%%%%%%%%%%%%%%%%%%%%%%%%%%%%%%%%%%%%%%%%%%%%%%%%%%%%%%%%%%%%%%%%%%
%%%%%%%%%%%%%%%%%%%%%%%%%%%%%%%%%%%%%%%%%%%%%%%%%%%%%%%%%%%%%%%%%%%%%%%%%%%%%%%%%%%%%%%%%%%%%%%%%%%%%%%%%%%%%%%%%%%%%%%%%%%%%%%%%%%%%%%%%%%%%%%%%
\appendix

\onecolumngrid

%%%%%%%%%%%%%%%%%%%%%%%%%%%%%%%%%%%%%%%%%%%%%%%%%%%%%%%%%%%%%%%%%%%%
%%%%%%%%%%%%%%%%%%%%%%%%%%%%%%%%%%%%%%%%%%%%%%%%%%%%%%%%%%%%%%%%%%%%
%%%%%%%%%%%%%%%%%%%%%%%%%%%%%%%%%%%%%%%%%%%%%%%%%%%%%%%%%%%%%%%%%%%%
%%%%%%%%%%%%%%%%%%%%%%%%%%%%%%%%%%%%%%%%%%%%%%%%%%%%%%%%%%%%%%%%%%%%
%%%%%%%%%%%%%%%%%%%%%%%%%%%%%%%%%%%%%%%%%%%%%%%%%%%%%%%%%%%%%%%%%%%%
\begin{figure*}[t!]
\subfigure[]{\includegraphics[width=0.22\linewidth]{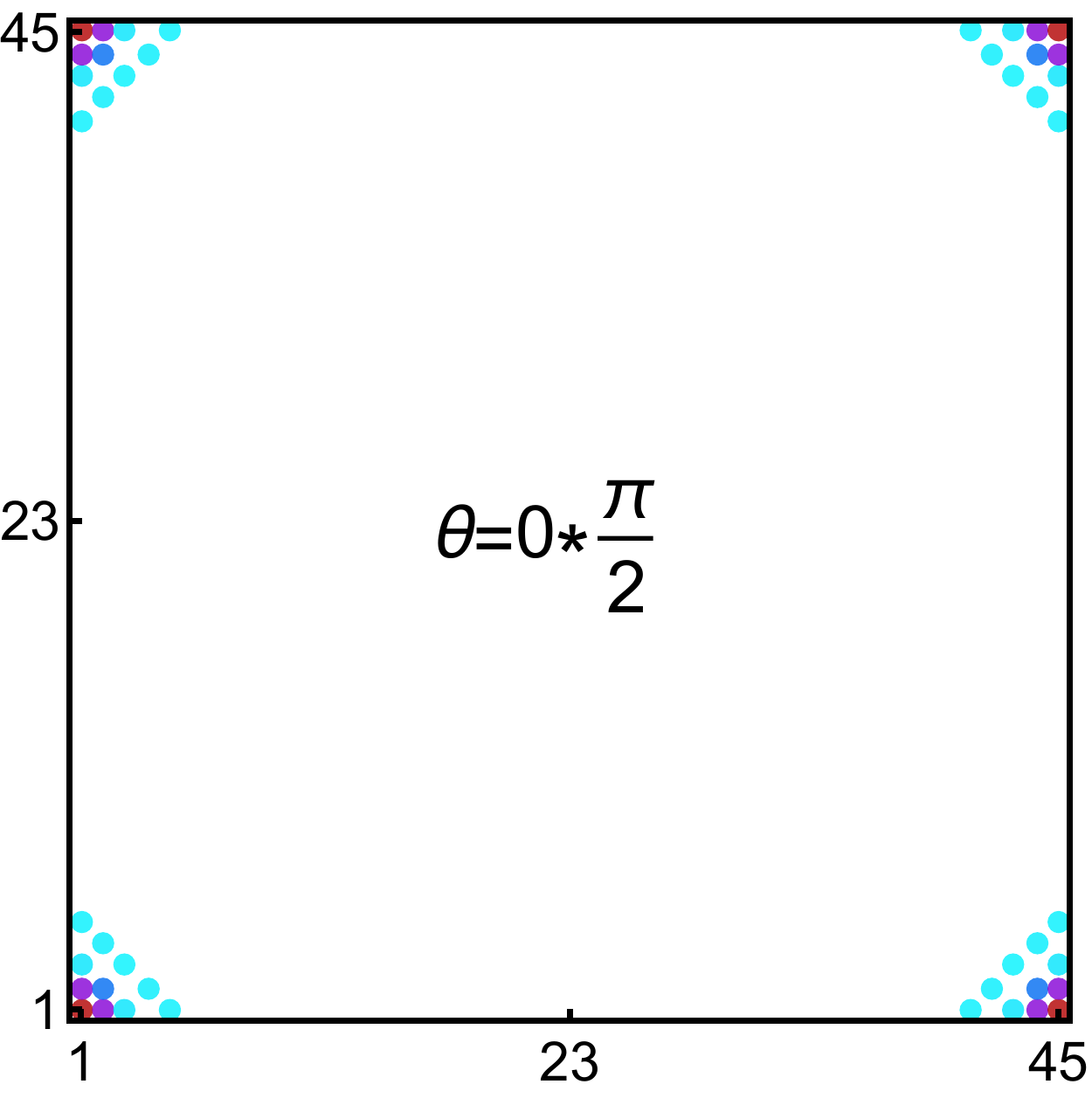}
\includegraphics[width=0.22\linewidth]{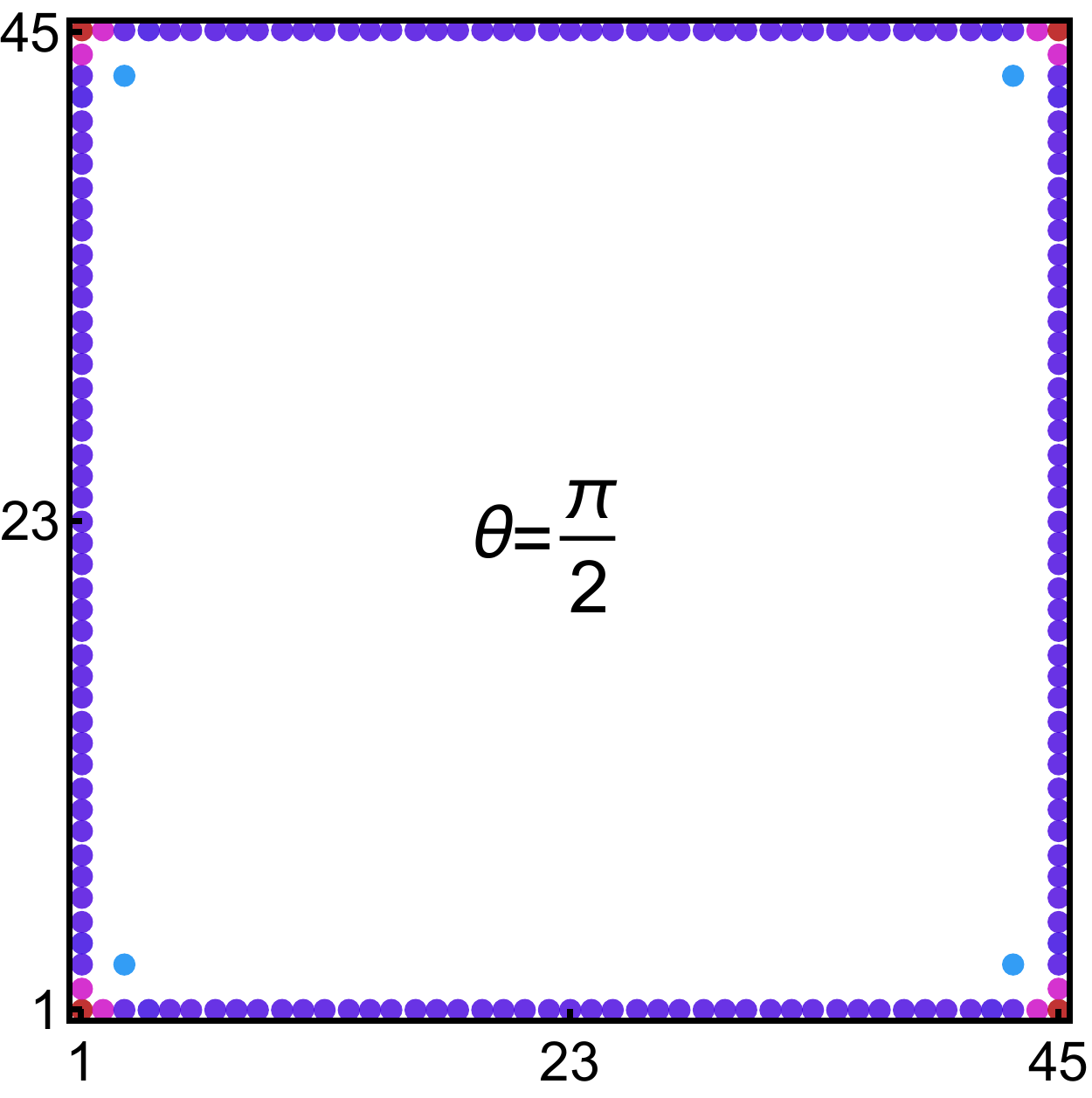}}%
\includegraphics[width=0.05\linewidth]{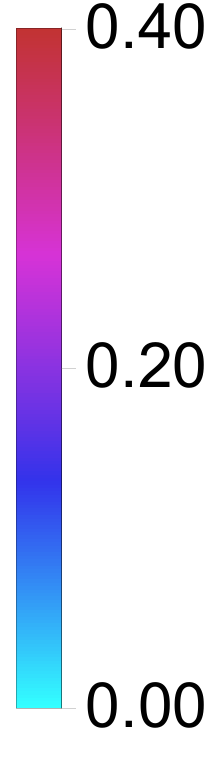}
\subfigure[]{\includegraphics[width=0.22\linewidth]{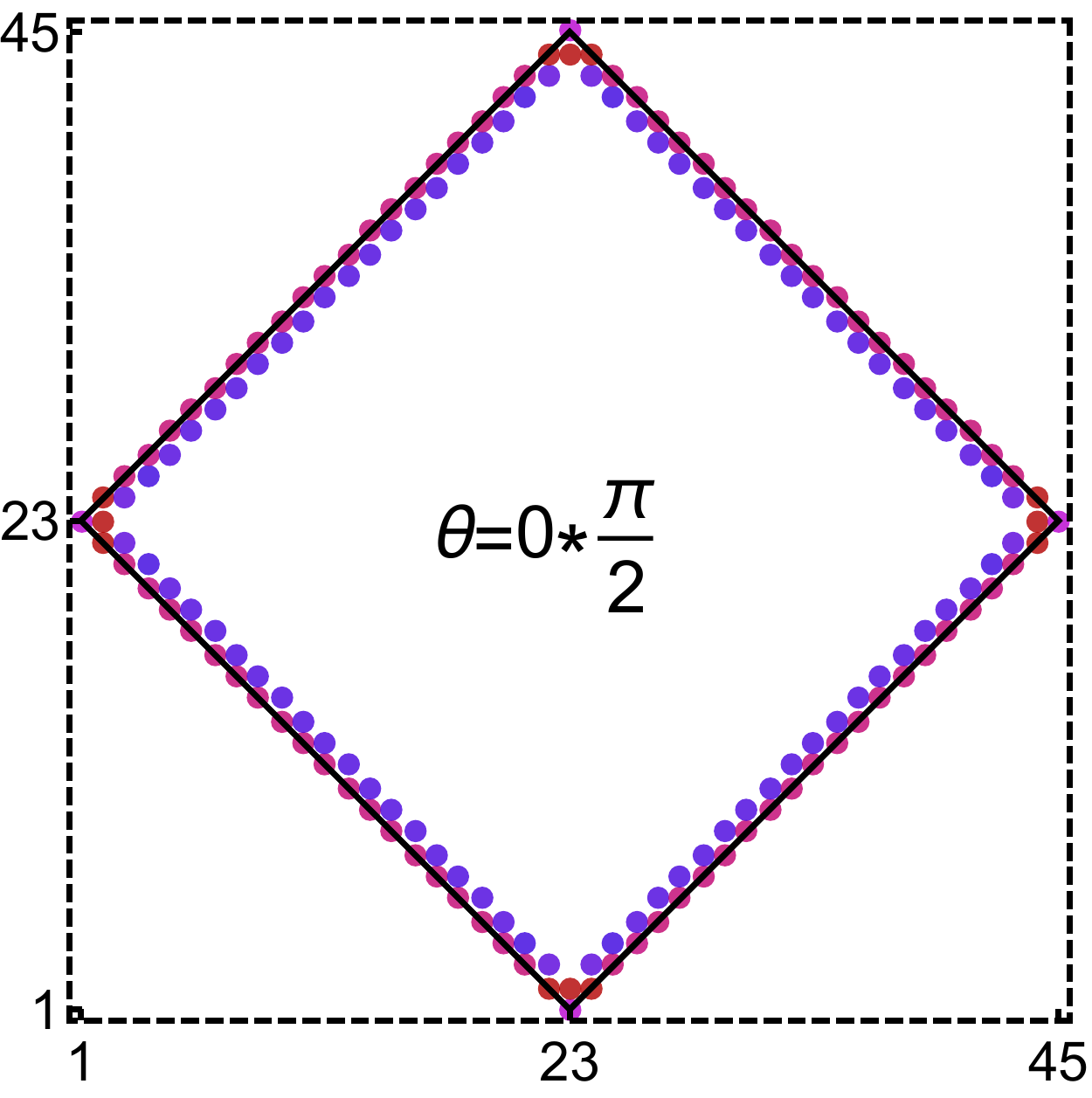}
\includegraphics[width=0.22\linewidth]{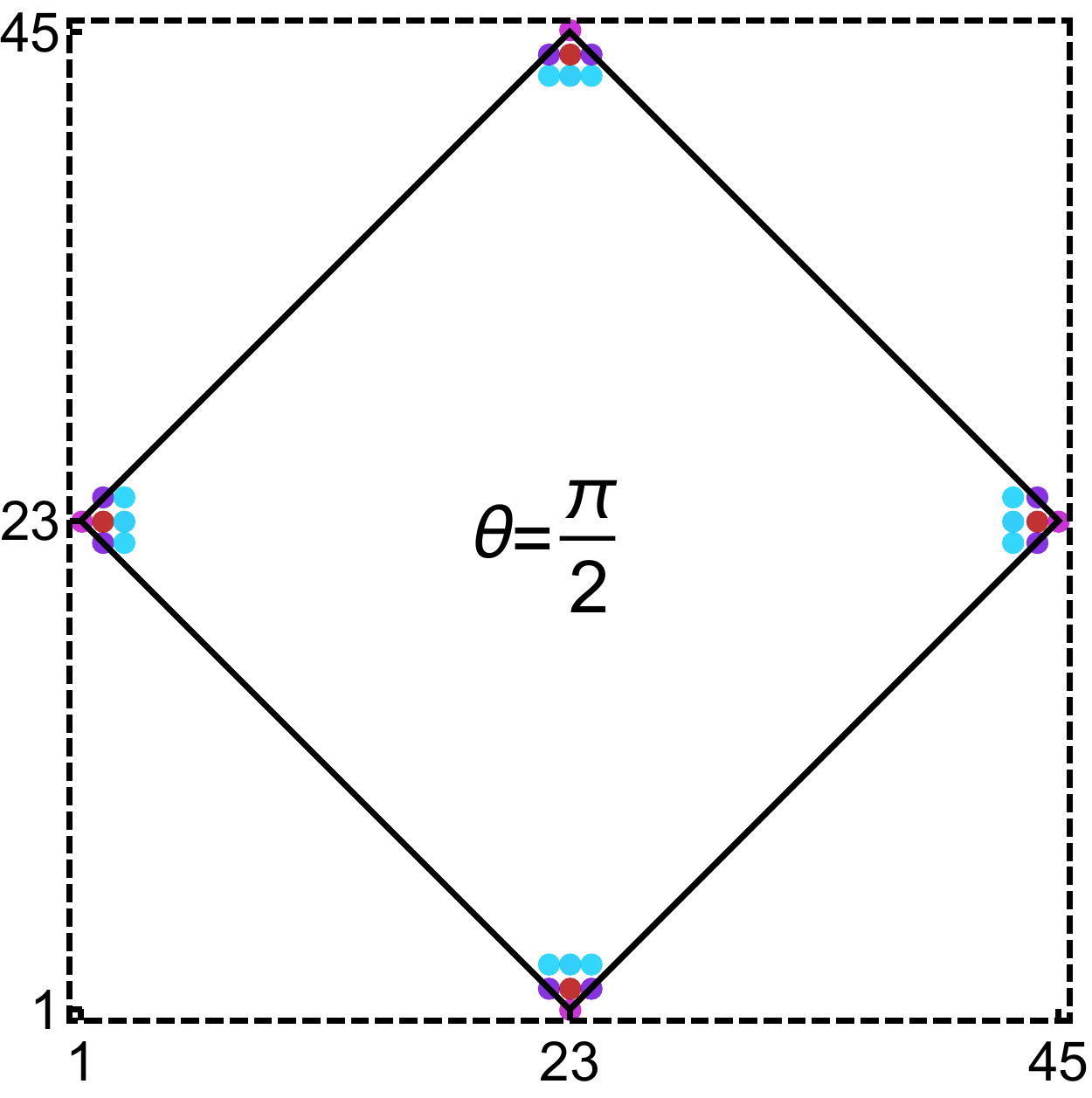}}%
\includegraphics[width=0.05\linewidth]{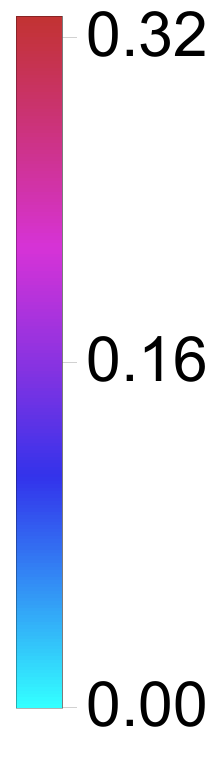}
\caption{Local density of states (LDoS) for four closest to the zero energy modes on (a) a regular square lattice (solid black box) and (b) an oblique square lattice (solid black box), cut from a regular square lattice (dashed black box). While the corners in (a) are placed along the body diagonal directions, they are placed along the principle axes in (b). The corresponding values of the parameter $\theta$ appearing in the second-order Wilson-Dirac mass are quoted in each figure [see Eq.~(\ref{eq:lattice-hamiltonian-HOT})]. Consequently, in panel (a) the corner modes appear when $\theta=0$, for which the HOT Wilson-Dirac mass changes its sign across the body diagonals, while the near zero energy modes occupy all the edges when $\theta=\pi/2$. On the other hand, in oblique square lattice in (b), the corner states appear when $\theta=\pi/2$ for which the second-order Wilson-Dirac mass changes its sign across the principle axes. The closet to zero energy states in this setup occupy all the edges when $\theta=0$.
}~\label{figappend:cornerevolution}
\end{figure*}
%%%%%%%%%%%%%%%%%%%%%%%%%%%%%%%%%%%%%%%%%%%%%%%%%%%%%%%%%%%%%%%%%%%%
%%%%%%%%%%%%%%%%%%%%%%%%%%%%%%%%%%%%%%%%%%%%%%%%%%%%%%%%%%%%%%%%%%%%
%%%%%%%%%%%%%%%%%%%%%%%%%%%%%%%%%%%%%%%%%%%%%%%%%%%%%%%%%%%%%%%%%%%%
%%%%%%%%%%%%%%%%%%%%%%%%%%%%%%%%%%%%%%%%%%%%%%%%%%%%%%%%%%%%%%%%%%%%
%%%%%%%%%%%%%%%%%%%%%%%%%%%%%%%%%%%%%%%%%%%%%%%%%%%%%%%%%%%%%%%%%%%%

\section{Dislocation modes in a 2D second-order $M$ phase: Continuum model}~\label{app:A}

In this Appendix we derive the form of the localized dislocation modes from the continuum model for the two-dimensional (2D) $M$ phase with a the Burgers vector ${\bf b}=a {\bf e}_x$. We start with the lattice Hamiltonian [Eqs. (1) and (2) in $d=2$ with specifically chosen form factors] $H=\sum_{{\bf k}}\Psi_{\bf k}^\dagger \hat{h} \Psi_{\bf k}$, where $\hat{h} = \hat{h}_{0}+\hat{h}_{\Delta}$, and
\begin{align}~\label{eq:lattice-hamiltonian-2DHOT}
\hat{h}_{0}&= t \left[ \sin(k_x a) \Gamma_1 + \sin(k_y a) \Gamma_2 \right]
+ \left\{m -2B \left[2- \cos(k_x a) - \cos(k_y a) \right] \right\} \Gamma_3, \nonumber\\
\hat{h}_{\Delta}&=\Delta \big\{\cos \theta \left[ \cos(k_x a) - \cos(k_y a) \right]
+ \sin \theta \left[ \sin(k_x a)\sin(k_y a) \right] \big\} \Gamma_4= \Delta({\bf k},\theta) \Gamma_4.
\end{align}
Here, $\Psi_{\bf k}$ is a four-component spinor, the exact form of which does not affect the following discussion, while ${\bf k}$ is the momentum, and $a$ is the lattice spacing. The mutually anticommuting four-component $\Gamma$ matrices satisfy the Clifford algebra $\{ \Gamma_j, \Gamma_k\}=2\delta_{jk}$ for $j,k=1, \cdots, 5$. The following discussion only rests on this anicommuting Clifford algebra, not on the exact representations of the $\Gamma$ matrices.
\\

This model for the regime of parameters $0<m/B<8$ describes a 2D first-order topological insulator and a HOT insulator (second-order) for $\Delta=0$ and finite $\Delta$, respectively. Furthermore, when $4< m/B < 8$, the model features the band inversion at the $M=(\pi/a,\pi/a)$ point (the $M$ phase), while for $0< m/B<4$, the band inversion is at the $\Gamma=(0,0)$ point (the $\Gamma$ phase) of the BZ. Notice that $\{\hat{h}_{0},\hat{h}_{\Delta}\}=0$, and therefore $\hat{h}_{\Delta}$ acts as a mass term for the topological edge states of $\hat{h}_0$. This mass term changes sign under the $C_4$ rotation and, as such, assumes the profile of a discrete symmetry breaking Wilson-Dirac mass, the exact form of which depends on the parameter $\theta\in[0,\pi/2]$. In particular, for $\theta=\pi/2$ the domain wall lies along the diagonals $k_y=\pm k_x$, while for $\theta=0$  it is located along the principal axes, $k_x=0, k_y=0$.
\\

We now comment on the structure of the corner modes for $\theta=0$ and $\pi/2$. Let us consider a 2D square lattice of linear dimension $L$ in each direction, such that four corners are at $(\pm L/2,\pm L/2)$. Four corner modes are then sharply localized around these corners for $\theta=0$, and with increasing $\theta$ they become more delocalized. By contrast, if the crystal is cut in such a way that four corners are located at $(\pm L/2,0)$ and $(0,\pm L/2)$, the corner modes are most prominently localized when $\theta=\pi/2$ and gradually delocalize as $\theta$ is ramped down to zero. However, irrespective of the sharpness of the corner modes, the system always describes a 2D second-order topological insulator for any $\theta$. These outcomes are shown in Fig.~\ref{figappend:cornerevolution}. A similar structure also appears for the hinge modes for 3D second-order topological insulator, discussed in Appendix~\ref{SMsec:3Ddislocation}.
\\

The continuum Hamiltonian is obtained by expanding the above lattice Hamiltonian  close to the bandgap closing at the $M$ point, $k_x=\pi+q_x$, $k_y=\pi+q_y$, with  $\Delta=0$, which in the real space (${\bf q}\rightarrow -i{\nabla}$) reads as \begin{equation}
{\hat h}_{\rm M}=it\Gamma_1 \partial_x+it\Gamma_2 \partial_y-\Gamma_3\left[{\tilde m}+{\tilde B}(\partial^2_x+\partial^2_y)\right].
\label{eq:Hamiltonian-M}
\end{equation}
Here, ${\tilde m}=8B-m>0$, ${\tilde B}=B>0$, and we set $a=1$.
\\

 The two edges  along the lines $x_\pm=\pm a$ [Fig.~1(a)], before the dislocation is introduced through the Volterra construction, feature topological gapless edge states, resulting from the corresponding zero modes of the edge Hamiltonian
$H_{\rm edge}=(it\Gamma_1\partial_x-({\tilde m}+{\tilde B}\partial^2_x)\Gamma_3) \otimes \mu_3$, where the vector of Pauli matrices ${\boldsymbol \mu}$ acts in the space of the two edges. When dislocation introduces a $\pi$ hopping phase, the reconnection of the edges across the Volterra cut is modeled by a hopping Hamiltonian between them in the form $H_{\rm D}= t\,{{\rm sgn} (x)}\Gamma_1\otimes \mu_1$, where ${{\rm sgn} (x)}$ is the ``sign'' function. This term, through the sign change of the hopping across the cut, takes into account that the (low-energy) electrons acquire a $\pi$ phase when encircling the defect. Its form ensures that when the phase factor is trivial, the connected edge modes are trivially gapped out, as in the $\Gamma$ phase. \\

We then look for the zero energy modes of the Hamiltonian $H_{\rm edge}+H_{\rm D}$, or explicitly
\begin{equation}\label{eq:explicit-zero-modes}
\left\{[it\Gamma_1\partial_x-({\tilde m}+{\tilde B}\partial^2_x)\Gamma_3]\otimes\mu_3
+t\,{{\rm sgn} (x)}\Gamma_1\otimes \mu_1\right\}\Psi_0(x)=0.
\end{equation}
After multiplying this equation from the left by $-i\Gamma_1\otimes\mu_3$,   we obtain
\begin{equation}\label{eq:explicit-zero-modes}
\left[it\partial_x+i\Gamma_1\Gamma_3({\tilde m}+{\tilde B}\partial^2_x)+t\,{{\rm sgn} (x)}\mu_2\right]\Psi_0(x)=0.
\end{equation}
The form of the above equation implies that we can take the following ansatz for the solution
\begin{equation}
\Psi_0(x)=f(x)\chi_\sigma\otimes \varphi_\rho,
\end{equation}
where
\begin{equation}
i\Gamma_1\Gamma_3\chi_\sigma=\sigma\chi_\sigma,\,\,
\mu_2 \varphi_\rho=\rho \varphi_\rho,
\label{eq:Dislocation-zero-modes}
\end{equation}
which for an exponentially localized solution at both edges $f(x)\sim \exp(-\lambda|x|)$, yields
\begin{equation}
-t\lambda+({\tilde m}+{\tilde B}\lambda^2)\sigma {\rm sgn} (x)+t\rho=0.
\end{equation}
Choosing $\sigma {\rm sgn} (x)=+1, \rho=+1$, and considering the regime $t^2>4{\tilde B}({\tilde m}+t)$, we obtain two characteristic  inverse localization lengths
\begin{equation}
\lambda_{1,2}=\frac{t\pm\sqrt{t^2-4{\tilde B}(t+{\tilde m})}}{2{\tilde B}}>0.
\end{equation}
Notice that the spinors $\chi_\sigma$ are doubly degenerate, because of the anticommuting property of the four-component Hermitian $\Gamma$ matrices. This form of the localization length together with the continuity condition across the reconnected edges, $\Psi_0(x\to0)=0$,
yields a pair of the zero energy dislocation modes  given by
\begin{equation}\label{eq:2D-modes}
\Psi_0^{(1,2)}(x)={\mathcal C}\chi^{(1,2)}_{{\rm sgn} (x)}\otimes\varphi_{+1}\left({\rm e}^{-\lambda_1|x|}-{\rm e}^{-\lambda_2|x|}\right)
\equiv {\mathcal C}\chi^{(1,2)}_{{\rm sgn} (x)}\otimes\varphi_{+1} \psi(x),
\end{equation}
mentioned in the main text. \\

The rest of the analysis straightforwardly follows from the fact that  $[ \Gamma_4,i\Gamma_1\Gamma_3 ]=0$ implying that $\Gamma_4$ reduces in the eigen-subspaces of $i\Gamma_1\Gamma_3$. Therefore, a perturbation proportional to $\Gamma_4$ symmetrically splits the dislocation modes about the zero energy. Explicitly, the HOT perturbation
\begin{equation}
\hat{h}_{\Delta}=\Delta \big\{\cos \theta \left[ \cos(k_x a) - \cos(k_y a) \right] + \sin \theta \sin(k_x a)\sin(k_y a) \big\} \Gamma_4,
\end{equation}
after the expansion about the $M$ point at ${\bf K}_{M}=(\pi/a,\pi/a)$, $k_x= \pi/a+q_x$, $k_y=\pi/a$, with $q_x=-i\partial_x$, takes the form
\begin{equation}
\hat{h}_{\Delta,M}=-\frac{1}{2}\Delta \cos \theta\partial_x^2.
\end{equation}
Therefore the splitting of the modes is given by
\begin{equation}
\delta E=2\Delta \cos\theta {\tilde E},
\end{equation}
where
\begin{equation}
{\tilde E}=\frac{1}{2} \left|\int dx \left( \Psi^{(1,2)}_0 (x) \right)^\dagger \partial_x^2 \Psi^{(1,2)}_0 (x)\right|,
\end{equation}
as announced in the main text. Finally, we note that in the above derivation of the dislocation modes in $d=2$ and also in $d=3$ (see  Appendix~\ref{SMsec:3Ddislocation}), we only used the Clifford algebra of the $\Gamma$ matrices, $\{\Gamma_i,\Gamma_j\}=2\delta_{ij}$, for $i,j=1,...,5$, and  the existence of the modes is therefore independent of the $\Gamma-$matrix representation. Since $[\Gamma_4, \Gamma_{13}]=0$, even though introduction of the HOT mass generically yields finite energy dislocation modes unless $\theta=\pi/2$, they cannot be mixed with rest of the bulk states, and therefore remain robust and protected. We further substantiate this claim by demonstrating the symmetry protection of the dislocation modes in Appendix~\ref{app:symmetry-protection} in the case of $C_{4n}$ rotational symmetry, while an analogous analysis for $C_{4n+2}$ rotational symmetry is presented  in  Appendix~\ref{app:F}.

%%%%%%%%%%%%%%%%%%%%%%%%%%%%%%%%%%%%%%%%%%%%%%%
%%%%%%%%%%%%%%%%%%%%%%%%%%%%%%%%%%%%%%%%%%%%%%%
%%%%%%%%%%%%%%%%%%%%%%%%%%%%%%%%%%%%%%%%%%%%%%%
%%%%%%%%%%%%%%%%%%%%%%%%%%%%%%%%%%%%%%%%%%%%%%%
%%%%%%%%%%%%%%%%%%%%%%%%%%%%%%%%%%%%%%%%%%%%%%%
\begin{figure*}[t!]
\subfigure[]{\includegraphics[width=0.32\linewidth]{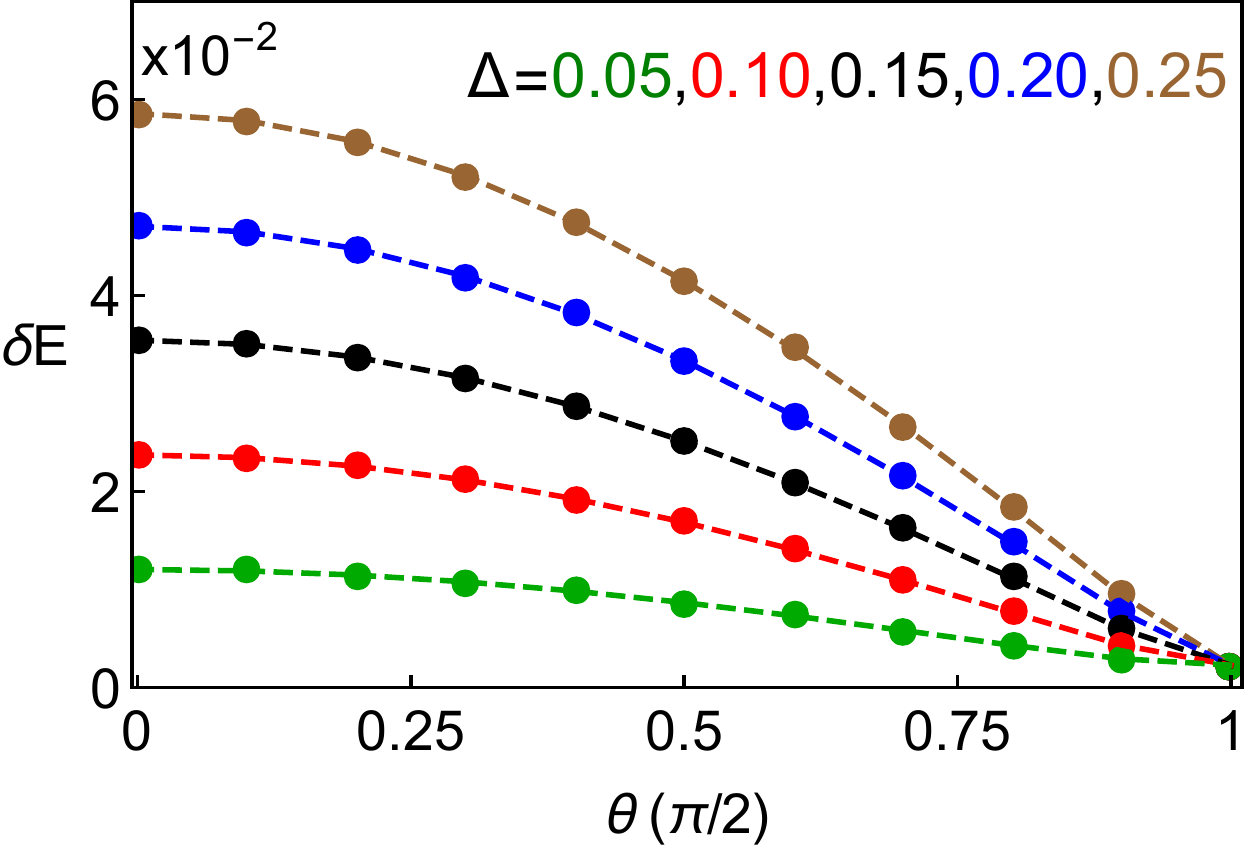}}
\subfigure[]{\includegraphics[width=0.32\linewidth]{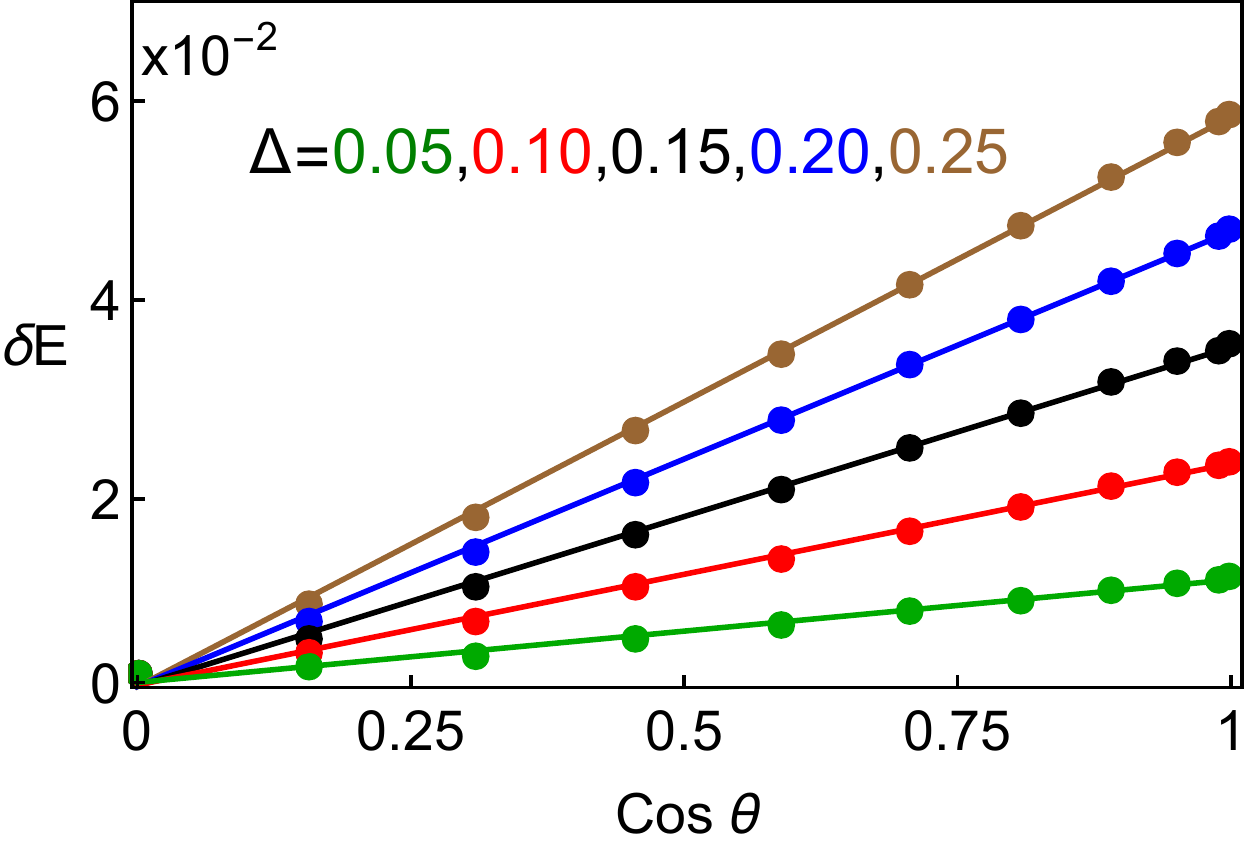}}
\subfigure[]{\includegraphics[width=0.32\linewidth]{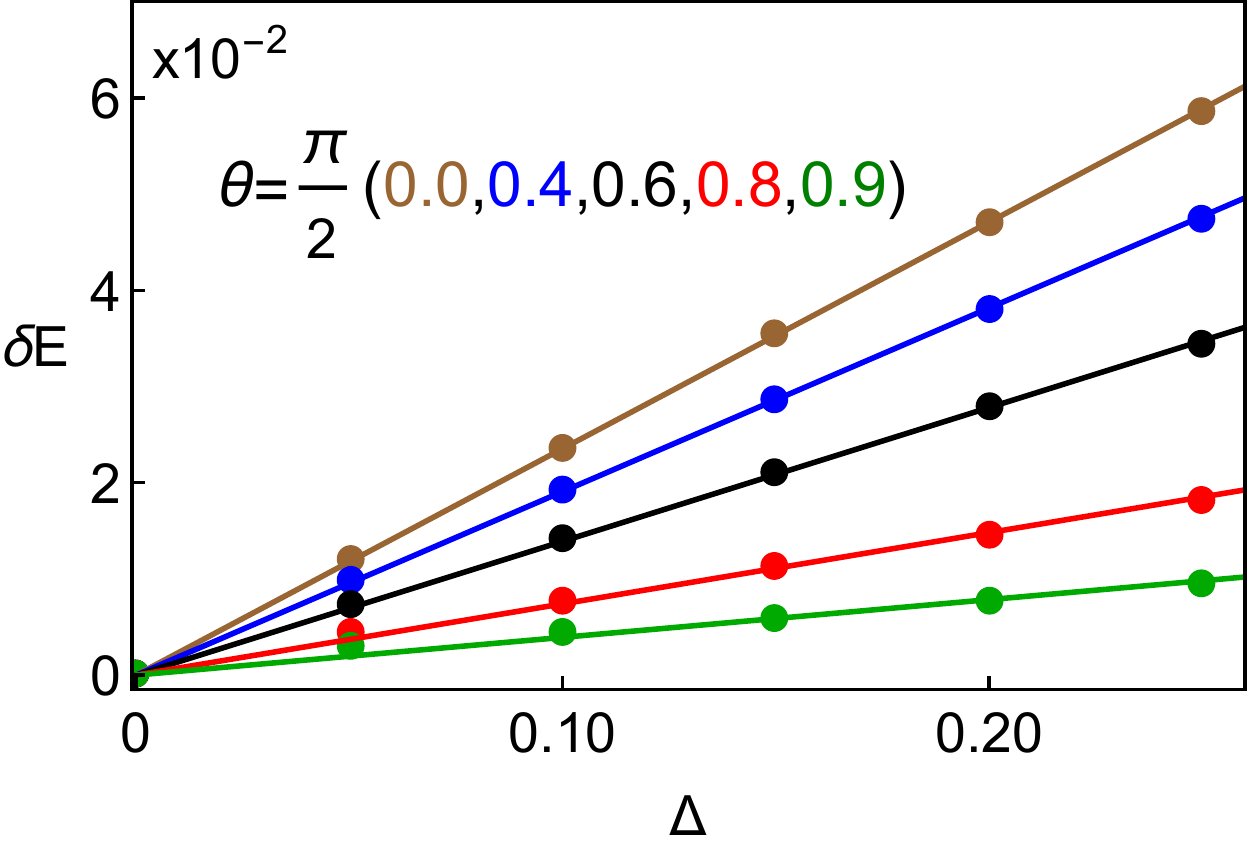}}
\subfigure[]{\includegraphics[width=0.28\linewidth]{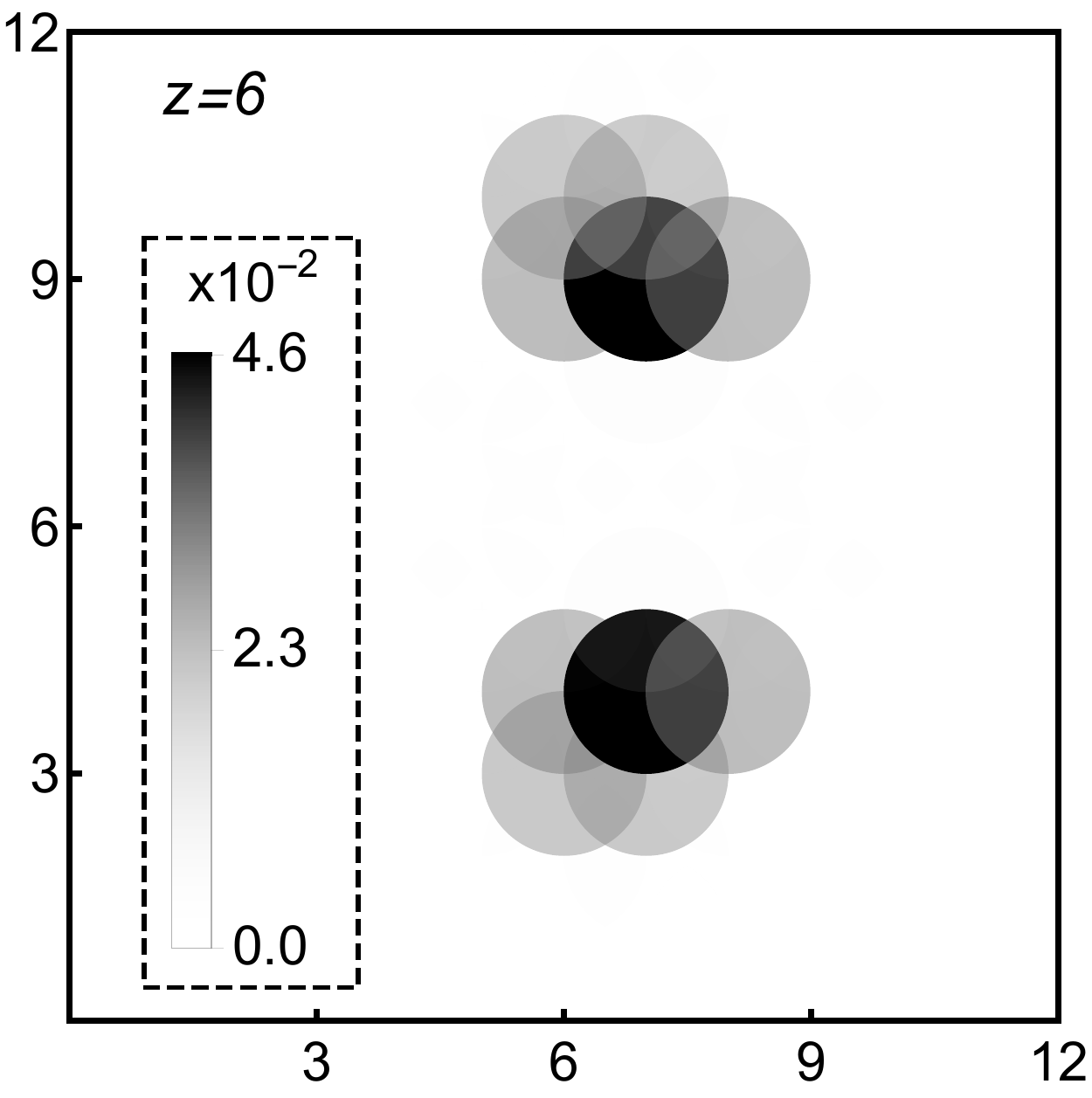}}
\subfigure[]{\includegraphics[width=0.28\linewidth]{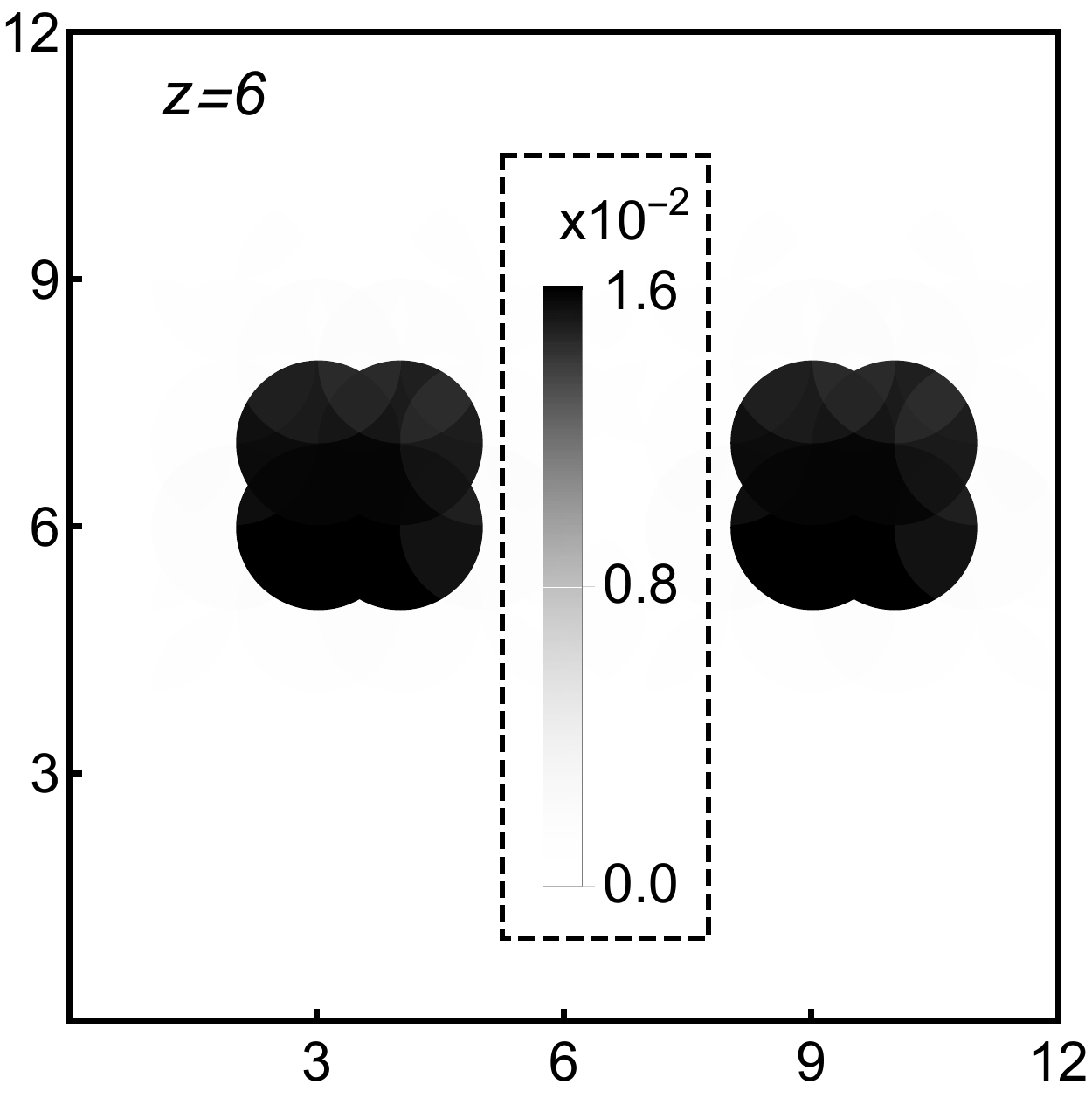}}
\subfigure[]{\includegraphics[width=0.38\linewidth]{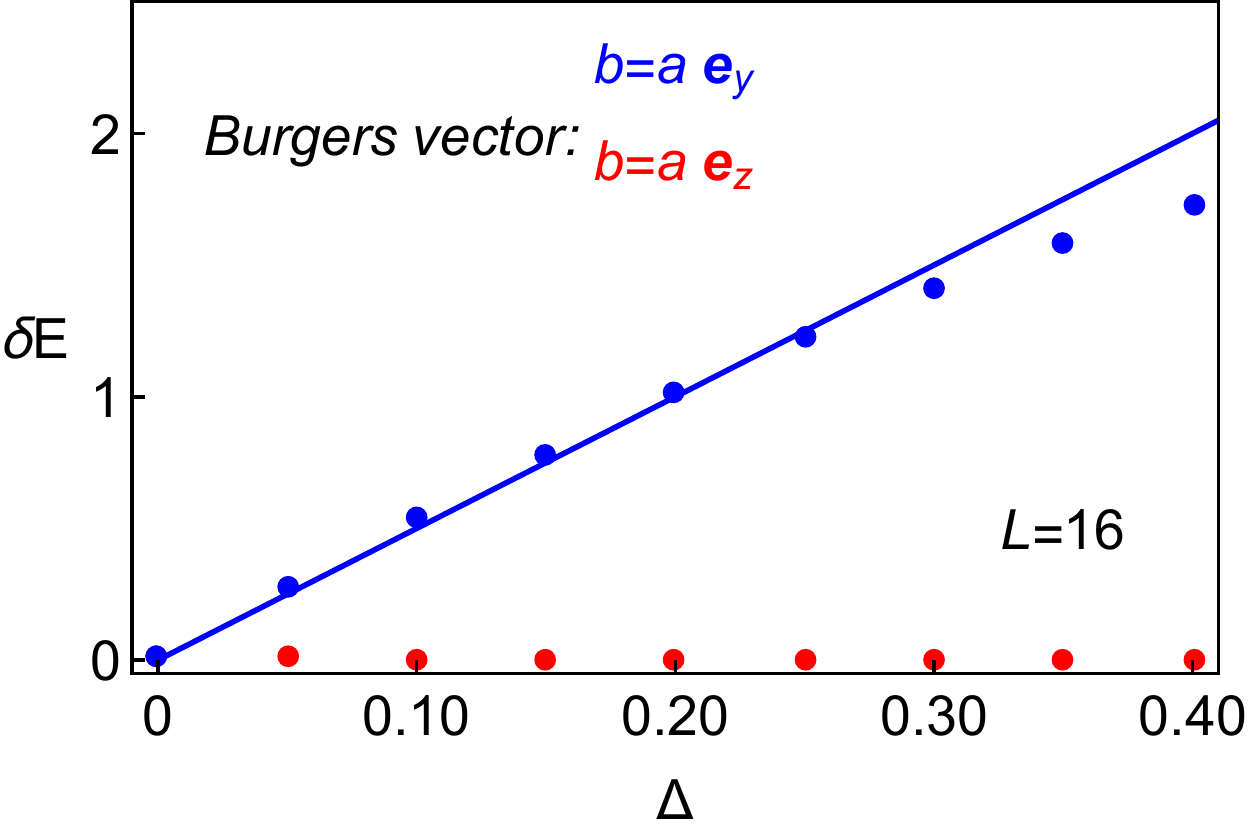}}
\caption{(a) Scaling of the spectral gap ($\delta E$) among four states localized at the core of the dislocations with the orientation of the $C_4$ symmetry breaking Wilson-Dirac mass $\theta$ [see Eq.~(\ref{eq:3DHOTHamltonian})] in a cubic system of linear dimension $L=16$ in each direction in the presence of an edge dislocation-antidislocation pair with the Burgers vector ${\bf b}=\pm a {\bf e}_x$. It shows that when the Burger vector is parallel to one of the $C_4$ symmetry breaking axis, i.e, $\theta=\pi/2$, edge dislocation modes become gapless. The linear scaling of the spectral gap $\delta E$ with (b) $\cos \theta$ and (c) $\Delta$, for small $\Delta$, are in agreement with our theoretical predictions from Appendix~\ref{SMsec:3Ddislocation}. Localization of the dislocation modes in the presence of (d) edge dislocation-antidislocation pair with the Burgers vector ${\bf b}=\pm a {\bf e}_x$, and (e) screw dislocation-antidislocation pair with the Burgers vector ${\bf b}=\pm a {\bf e}_z$, on a specific $xy$ plane along the $z$ direction ($z=6$) in a cubic system with linear dimension $L=12$ in each direction. The direction of the HOT mass domain wall is fixed to be $\theta=0$ [Eq.~\eqref{eq:3DHOTHamltonian}]. It shows that these modes are always highly localized around the core of dislocations. (f) Scaling of the spectral gap ($\delta E$) among the states localized at the dislocation core in a system of linear dimension $L=16$ in each direction, in the presence of a screw dislocation-antidislocation pair, when the Burgers vector is ${\bf b}=\pm a {\bf e}_z$ (red) and ${\bf b}=\pm a {\bf e}_y$ (blue). Respectively in these two cases the screw dislocation pierces surfaces hosting gapless and gapped modes, and concomitantly the dislocation modes are also gapless and gapped.
}~\label{FigSM:3Ddislocation}
\end{figure*}
%%%%%%%%%%%%%%%%%%%%%%%%%%%%%%%%%%%%%%%%%%%%%%%
%%%%%%%%%%%%%%%%%%%%%%%%%%%%%%%%%%%%%%%%%%%%%%%
%%%%%%%%%%%%%%%%%%%%%%%%%%%%%%%%%%%%%%%%%%%%%%%
%%%%%%%%%%%%%%%%%%%%%%%%%%%%%%%%%%%%%%%%%%%%%%%
%%%%%%%%%%%%%%%%%%%%%%%%%%%%%%%%%%%%%%%%%%%%%%%

\section{3D edge and screw dislocation modes: The continuum model}~\label{SMsec:3Ddislocation}

In this Appendix we start with the lattice model for a three-dimensional second-order topological insulator on a cubic lattice, analogous to Eq.~\eqref{eq:lattice-hamiltonian-2DHOT}, but with [Eqs.~(1) and~(2) in $d=3$]
\begin{align}
h_{0}&= t \left[ \sin(k_x a) \Gamma_1 + \sin(k_y a) \Gamma_2 +\sin(k_z a) \Gamma_3 \right]
+ \left\{m - 2B \left[ 3-\cos(k_x a) - \cos(k_y a)-\cos(k_z a) \right] \right\} \Gamma_4, \nonumber\\
\hat{h}_{\Delta}&=\Delta \big\{\cos \theta \left[ \cos(k_x a) - \cos(k_y a) \right]
+ \sin \theta \left[ \sin(k_x a)\sin(k_y a) \right] \big\} \Gamma_5 = \Delta({\bf k}, \theta) \Gamma_5.
\label{eq:3DHOTHamltonian}
\end{align}
We now consider the $R$ phase, with the band inversion at the $R=(\pi/a,\pi/a,\pi/a)$ point in the BZ, realized for the values of the parameters $8<m<12$, $t=1$ and $B=1$, and expand the above Hamiltonian about the $R$ point for $\Delta=0$ to obtain
\begin{equation}\label{eq:cont-3DHam}
h_{0,R}(k_x,k_y,k_z)= t \left[ -k_x \Gamma_1 - k_y \Gamma_2 -k_z \Gamma_3 \right] + (-{\tilde m}+Bk^2) \Gamma_4,
\end{equation}
where ${\tilde m}=12B-m>0$ in the topological $R$ phase. We take $B=1$, and $k_i=\pi/a+q_i$, for $i=x,y,z$.  \\

A 3D edge dislocation is obtained by translating its 2D counterpart along a particular lattice direction representing the defect line. Therefore, the conclusions obtained above in the 2D case for a Burgers vector along the $C_4$ symmetry breaking $x$ or $y$ direction  directly apply to the 3D case: an edge dislocation should in general feature gapped modes which become gapless when the Burgers vector is parallel to the domain wall ($\theta=\pi/2$), consistent with our numerical findings, see Fig.~\ref{FigSM:3Ddislocation}(a)-(c). Also, when the Burgers vector is oriented along the $C_4$ symmetry axis, therefore piercing a gapless surface (in this case the $xy$ surface), the defect should feature gapless modes.
\\

A screw dislocation with the Burgers vector ${\bf b}=a{\bf e}_z$ can be introduced by the Volterra construction as follows.
We first choose a slip half-plane for $y=0$, $x>0$, whose neighbor half-plane $y=a{\bf e}_y$, $x>0$ is displaced by the Burgers vector
${\bf b}=a{\bf e}_z$ relative to the slip plane. Translational symmetry is then restored by reconnecting the bonds
between the slip and the neighbor plane everywhere, except close to the dislocation line along the $z$-direction.
An electron sliding down through the dislocation picks up a phase $\Phi_{\rm dis}={\bf K}_{\rm inv}\cdot{\bf b}=\pi$ upon encircling it once.
\\

The Hamiltonian for the slip plane and its neighbor before the reconnection reads
\begin{equation}\label{eq:3D-surfaceHamiltonian}
h_{xz}(k_x,k_z)=h_{0,R}(k_x,\pi/a,k_z)\otimes\mu_3,
\end{equation}
where the vector of Pauli matrices ${\bm \mu}$ acts in the space of the two surfaces.
The $\pi$ phase factor is then introduced by modifying the sign of the hopping in the $y$-direction across the domain wall at $x=0$, because the slip plane is orthogonal to the $y$-axis, yielding
\begin{equation}\label{eq:3D-hopping-modification}
h_{\rm 3D-dis}=t\,{\rm sgn} (x) \; \Gamma_1\otimes\mu_1.
\end{equation}
The form of this Hamiltonian is chosen so that when the phase factor is trivial, the surface zero modes are gapped out, as in the $\Gamma$ phase, realized for $t=B=1$ and $0<m<4$ in Eq.~\eqref{eq:3DHOTHamltonian}.
We now look for the zero modes of the surface Hamiltonian using that the translational symmetry is preserved
in the $z$-direction and is explicitly broken in the $x$-direction. Hence, $k_x=\pi/a-i\partial_x$, $k_z=\pi/a$, and
\begin{equation}
\left\{[it\Gamma_1\partial_x-({\tilde m}+{\tilde B}\partial^2_x)\Gamma_4]\otimes\mu_3
+t\,{{\rm sgn} (x)}\Gamma_1\otimes \mu_1\right\}\Psi_{0,{\rm 3D}}(x)=0,
\end{equation}
which is identical to the Eq.~\eqref{eq:explicit-zero-modes}, up to the change $\Gamma_3\rightarrow\Gamma_4$. The form of the pair of zero modes is therefore given by Eq.~\eqref{eq:2D-modes}, with the difference that here
$i\Gamma_1\Gamma_4\chi_\sigma=\sigma\chi_\sigma$. Gapless propagating modes along the dislocation line are obtained by ``translating" the zero modes along the dislocation direction $\Psi^{(1,2)}_{\rm 3D}(k_z)\sim\Psi^{(1,2)}_{0,{\rm 3D}}(x)e^{ik_z z}$. Notice that this solution
explicitly breaks $C_4$ symmetry which is a consequence of the specific Volterra cut not preserving this symmetry. Namely, there are two equivalent slip planes, $x-z$ and $y-z$,  each of them individually breaking $C_4$ symmetry.
On the lattice, this symmetry is certainly preserved away from the dislocation line
and therefore the zero mode solution in the continuum inherits it, so that, after a proper regularization
the spatially dependent part should obey  $\Psi_{\rm 3D}(x,y)=\Psi_{\rm 3D}(-y,x)$. This feature, as we show below is inherited by the HOT dislocation modes, and as we also find in our numerical analysis, see Fig.~\ref{FigSM:3Ddislocation}(e) for the plot of the local density of states (LDoS) of the closest to zero energy modes on a particular $x-y$ plane.
\\

%%%%%%%%%%%%%%%%%%%%%%%%%%%%%%%%%%%%%%%%%%%%%%%
%%%%%%%%%%%%%%%%%%%%%%%%%%%%%%%%%%%%%%%%%%%%%%%
%%%%%%%%%%%%%%%%%%%%%%%%%%%%%%%%%%%%%%%%%%%%%%%
%%%%%%%%%%%%%%%%%%%%%%%%%%%%%%%%%%%%%%%%%%%%%%%
%%%%%%%%%%%%%%%%%%%%%%%%%%%%%%%%%%%%%%%%%%%%%%%
\begin{figure*}[t!]
\subfigure[]{\includegraphics[width=0.48\linewidth]{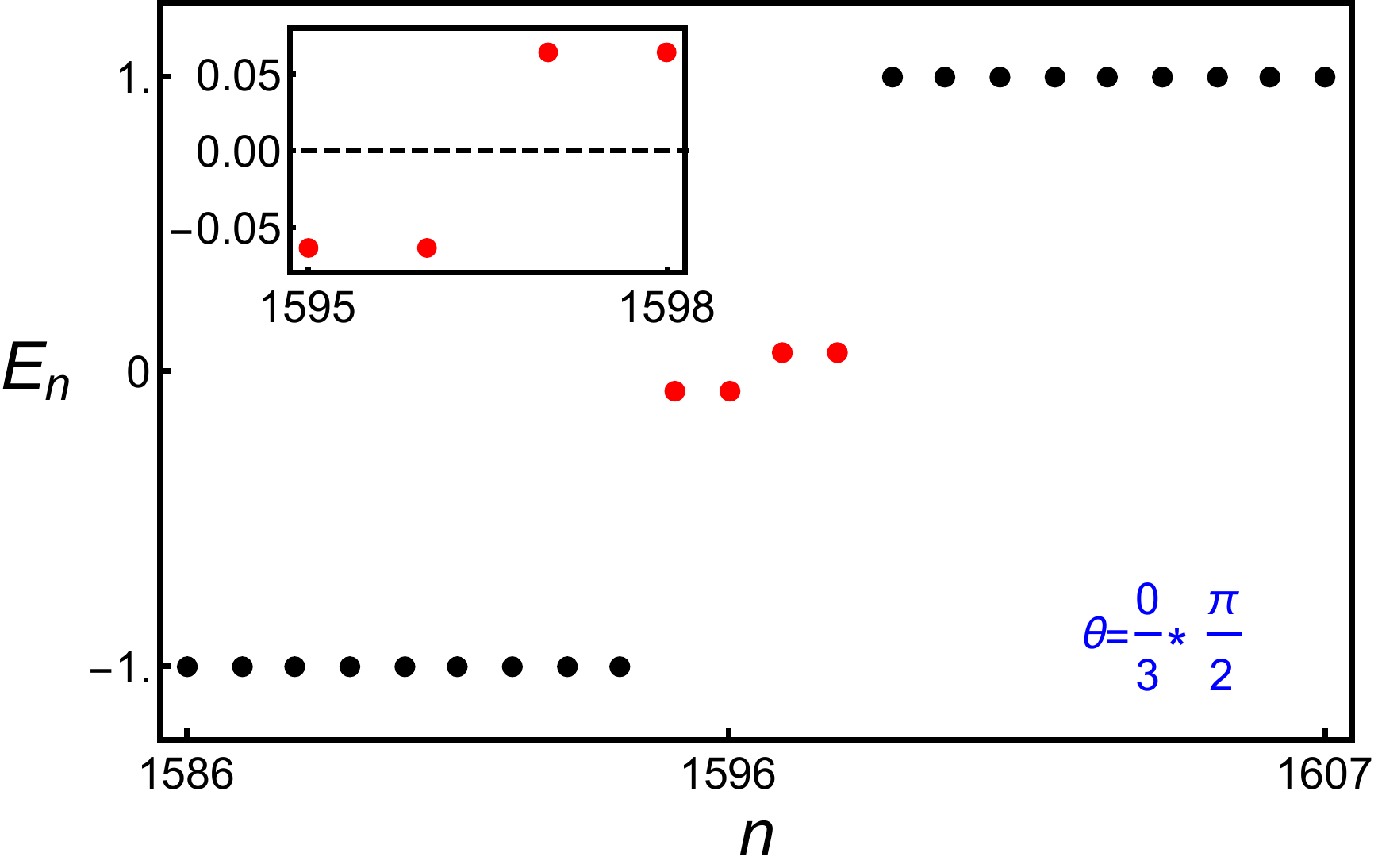}}%
\subfigure[]{\includegraphics[width=0.49\linewidth]{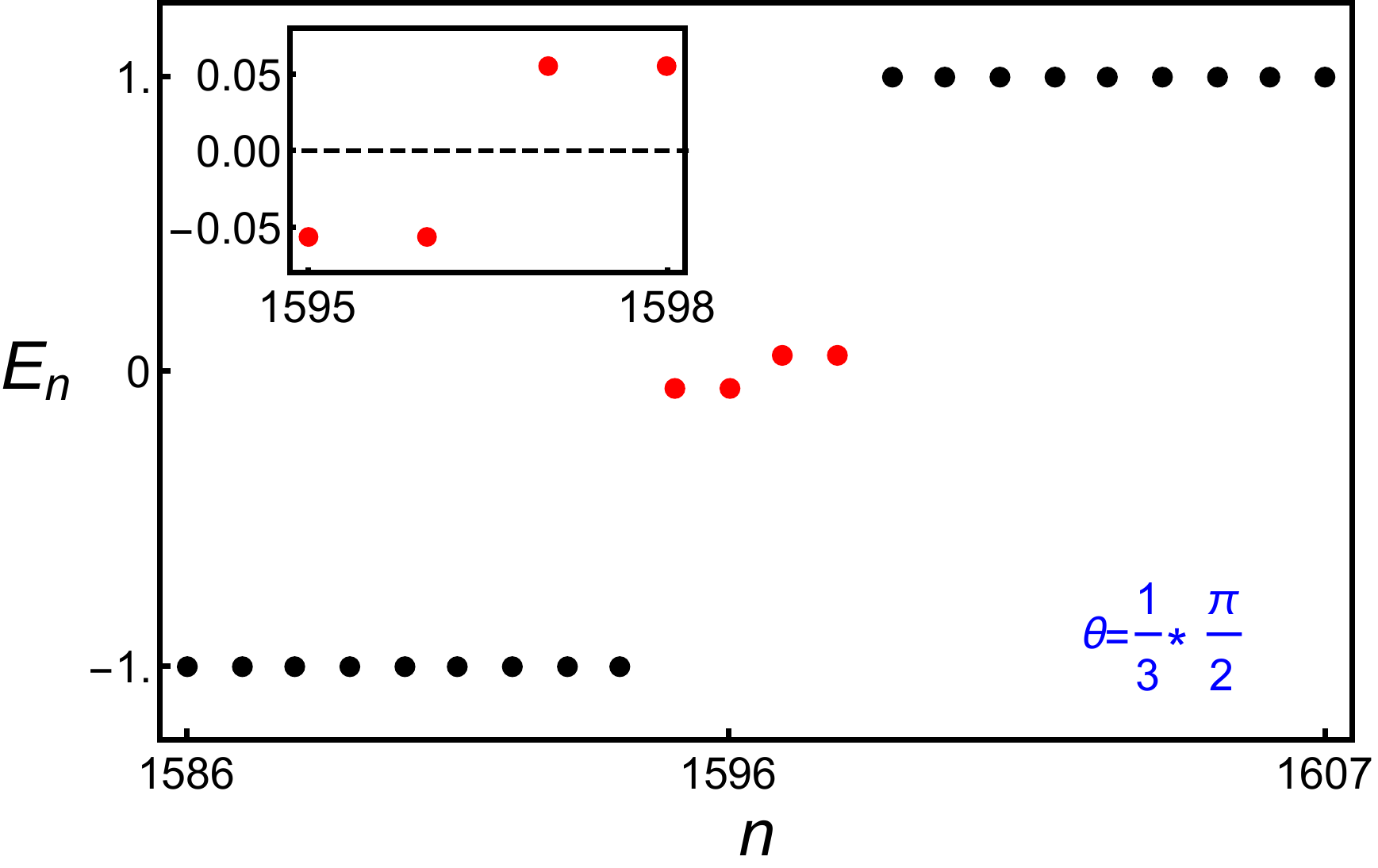}}
\subfigure[]{\includegraphics[width=0.49\linewidth]{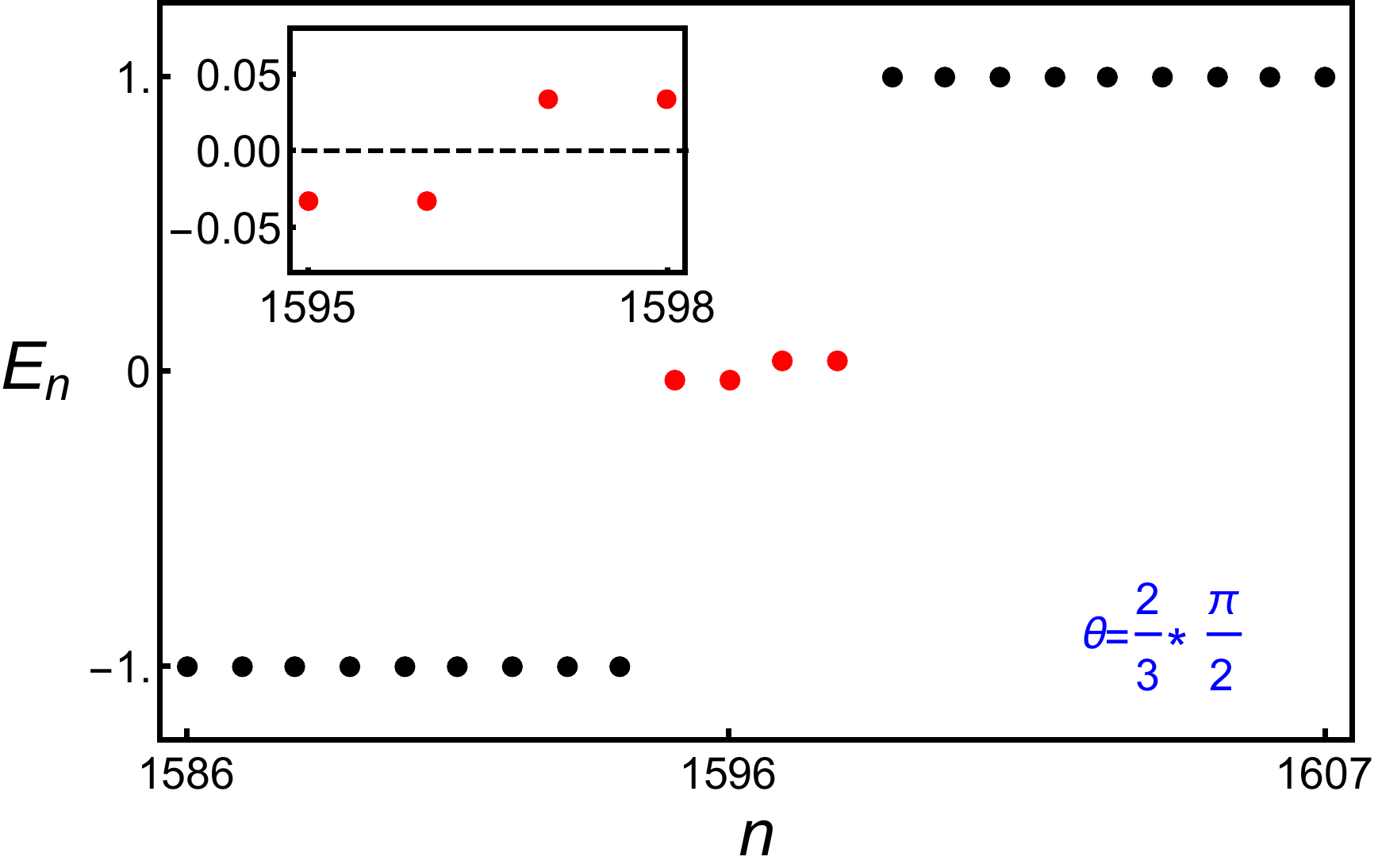}}%
\subfigure[]{\includegraphics[width=0.49\linewidth]{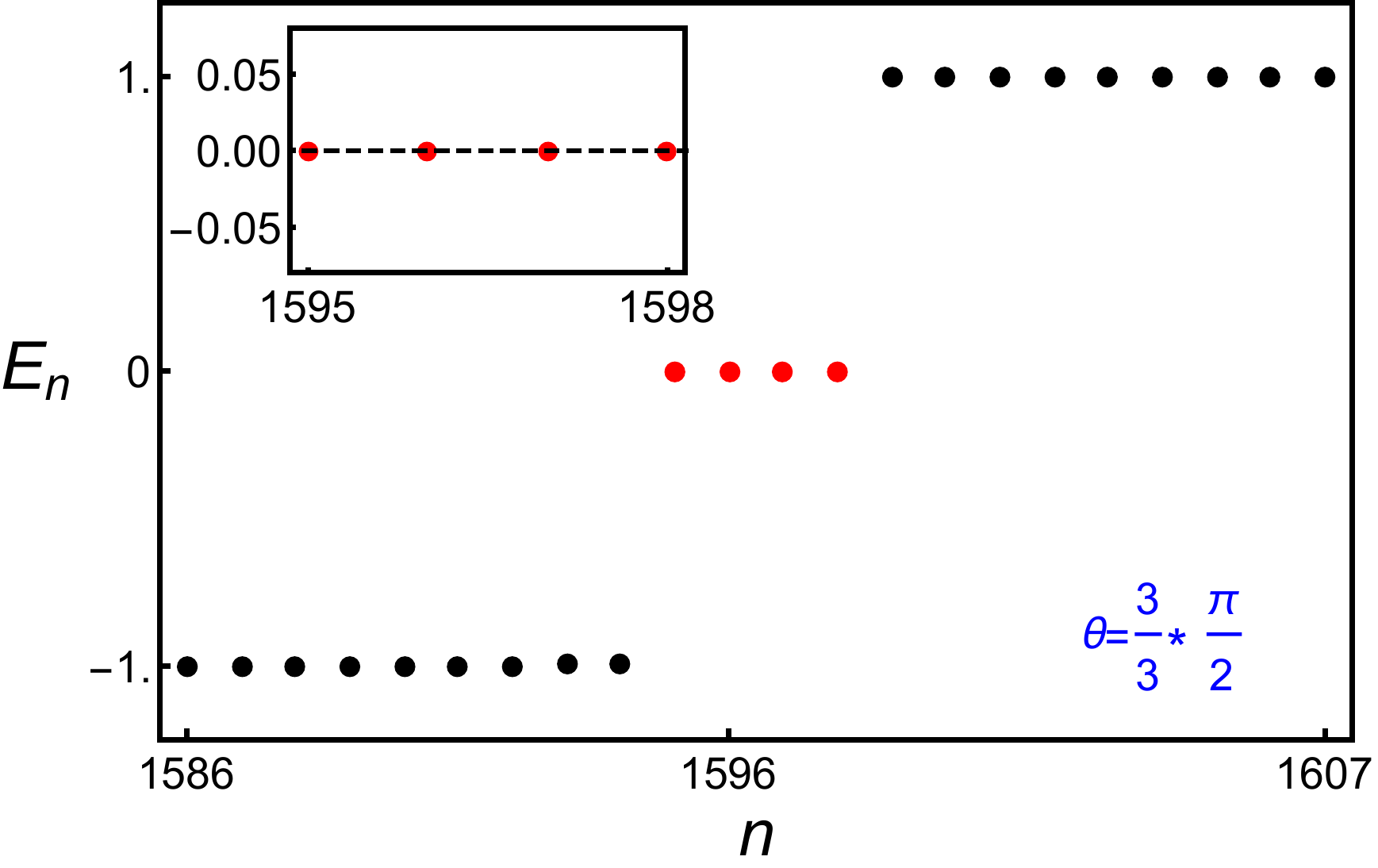}}
\caption{Spectral flow of the dislocation modes (red) in a system of linear dimension $L=28$ and with the periodic boundary in both directions in a two-dimensional second-order topological insulator with a pair of edge dislocation-antidislocation, for $t=2B=1.0$, $m=3.0$ and $\Delta=0.20$, as a function of the relative angle ($\theta$) between the Burgers vector ${\bf b}=\pm a {\bf e}_x$ and the HOT mass domain wall (see Fig.~1(a) of the main text). The bulk states are shown in black. As the HOT mass domain wall becomes parallel to the Burgers vector ($\theta=\pi/2$), the dislocation modes become zero-energy states. Note that for any value of $\theta$ the dislocation modes (at zero energy or finite energies) are always well separated from the bulk states, and they are particle-hole partners of each other, due to the unitary particle-hole symmetry generated by $\Theta=\Gamma_5$ (see Appendix~\ref{secSI:particlehole}), and thus do not mix with the bulk states. Hence, they are protected by the bulk topology and cannot be removed from the system (by mixing with the bulk states, for example). Here, $n$ is the index for the energy eigenvalues ($E_n$). In Appendix~\ref{app:symmetry-protection}, we also show that the composite symmetries of HOT insulators forbid addition of any new perturbation in the system, leaving the dislocation modes at finite and zero energy symmetry protected.
}~\label{FigSM:soectralflow2D}
\end{figure*}
%%%%%%%%%%%%%%%%%%%%%%%%%%%%%%%%%%%%%%%%%%%%%%%
%%%%%%%%%%%%%%%%%%%%%%%%%%%%%%%%%%%%%%%%%%%%%%%
%%%%%%%%%%%%%%%%%%%%%%%%%%%%%%%%%%%%%%%%%%%%%%%
%%%%%%%%%%%%%%%%%%%%%%%%%%%%%%%%%%%%%%%%%%%%%%%
%%%%%%%%%%%%%%%%%%%%%%%%%%%%%%%%%%%%%%%%%%%%%%%

We now introduce the HOT mass term proportional to $\Gamma_5$, as given by the Hamiltonian $h_\Delta$ in  Eq.~\eqref{eq:3DHOTHamltonian}, that may gap out otherwise gapless propagating modes along the dislocation since $[\Gamma_5,i\Gamma_1\Gamma_4]=0$, analogous as in the case in two dimensions. The splitting depends also on the form factor $\Delta({\bf k}, \theta)$, which after expanding about the band inversion $R$ point becomes
\begin{equation}
\Delta(x,y)=\frac{\Delta}{2}\left\{\cos\theta(\partial_y^2-\partial_x^2)+2 \sin \theta \partial_x\partial_y\right\},
\end{equation}
which then yields for the energy gap
\begin{equation}
\delta E\sim \int dx dy \Psi^*(x,y) \{ \cos\theta(\partial_y^2-\partial_x^2) + 2 \sin\theta \partial_x \partial_y\} \Psi(x,y)=0,
\end{equation}
because of the $C_4$ symmetry [both $(\partial_x^2-\partial_y^2)$ and $\partial_x\partial_y$ are odd, while $\Psi(x,y)$ is even under the $C_4$]. The $x-y$ surface therefore yields gapless modes
for the screw dislocation piercing it, as also expected from the fact that it is gapless. This is precisely what we find in the numerical analysis, see Fig.~\ref{FigSM:3Ddislocation}(f) (red dots). The plot of the LDoS of the closest to zero energy modes shows explicitly the $C_4$ symmetry, see Fig.~\ref{FigSM:3Ddislocation}(e). The result can also be appreciated from the fact that
in the continuum limit, featuring the full $U(1)$ rotational symmetry in the $x-y$ plane,
the dislocation represents a $\pi$ magnetic flux tube, which is known to carry the gapless mode when piercing a gapless surface~\cite{ran-natphys2009, slager-prb2014}.
On the other hand, when the screw dislocation is oriented in the $x$ or $y$  direction, piercing therefore a gapped surface,
it carries gapped modes.  This can be directly seen by
recalling that the dislocation zero modes have to break the $C_4$ symmetry, and thus the spatial part of the zero energy
solution for the dislocation with Burgers vector in the $y$ direction is $\Psi(x)$, as given in Eq.~\eqref{eq:2D-modes}.
The energy splitting is then
\begin{equation}
\delta E \sim \Delta \cos\theta \int dx \psi(x)\partial_x^2\psi(x) \neq 0,
\end{equation}
for any $\theta \neq \pi/2$. This is consistent with the numerically extracted scaling of $\delta E$ versus the HOT mass parameter $\Delta$, shown in Fig.~\ref{FigSM:3Ddislocation}(f) (blue dots).
Therefore, as long as the screw dislocation pierces the gapped surfaces, the modes along the dislocation line are also gapped with the gap scaling with the size of the surface gap ($\Delta$). The dislocation modes, however, remain within the bulk bandgap, determined by the first-order topological mass proportional to the $\Gamma_4$ matrix in Eq.~(\ref{eq:3DHOTHamltonian}).

%%%%%%%%%%%%%%%%%%%%%%%%%%%%%%%%%%%%%%%%%%%%%%%%%%%%%%%%%%%%%%%%%%%%%%%%%%%%%%%%%%%%%%%%%%%%%%%%%%%%%%%%%%%%%%%%%%%%
%%%%%%%%%%%%%%%%%%%%%%%%%%%%%%%%%%%%%%%%%%%%%%%%%%%%%%%%%%%%%%%%%%%%%%%%%%%%%%%%%%%%%%%%%%%%%%%%%%%%%%%%%%%%%%%%%%%%
%%%%%%%%%%%%%%%%%%%%%%%%%%%%%%%%%%%%%%%%%%%%%%%%%%%%%%%%%%%%%%%%%%%%%%%%%%%%%%%%%%%%%%%%%%%%%%%%%%%%%%%%%%%%%%%%%%%%
%%%%%%%%%%%%%%%%%%%%%%%%%%%%%%%%%%%%%%%%%%%%%%%%%%%%%%%%%%%%%%%%%%%%%%%%%%%%%%%%%%%%%%%%%%%%%%%%%%%%%%%%%%%%%%%%%%%%
%%%%%%%%%%%%%%%%%%%%%%%%%%%%%%%%%%%%%%%%%%%%%%%%%%%%%%%%%%%%%%%%%%%%%%%%%%%%%%%%%%%%%%%%%%%%%%%%%%%%%%%%%%%%%%%%%%%%
\begin{table}
    \begin{tabular}{|c|c|c|c|c|c|c|}
    \hline
    \multirow{2}{*}{Operator}& \multicolumn{6}{|c|}{Symmetry}\\
    \cline{2-7}
     & ${\mathcal T}$ & ${\mathcal P}$ & $C_4$& ${\mathcal P}{\mathcal T}$ & $C_4{\mathcal P}$ & $C_4{\mathcal T}$ \\
    \hline
    $\hat{h}_{0}$  & $\checkmark$ & $\checkmark$ & $\checkmark$ & $\checkmark$ & $\checkmark$ & $\checkmark$\\ \hline
    $\hat{h}_{\Delta}$  & $\times$ & $\times$ & $\times$ & $\checkmark$ & $\checkmark$ & $\checkmark$\\ \hline
    \end{tabular}
    \caption{Symmetry transformations of the first-order Hamiltonian ($\hat{h}_{0}$) and the second-order mass term ($\hat{h}_{\Delta}$), given by Eq.~(1) in the main text. The representation  of the $\Gamma$ matrices is given in Eq.~\eqref{eq:gammac4breaking}. Time-reversal (${\mathcal T}$),  parity (${\mathcal P}$) and $C_4$ rotational (${\mathcal R}_4$) symmetries are, respectively, represented by the operators ${\mathcal T}=\tau_2\otimes\sigma_1 {\mathcal K}$, ${\mathcal P}=\tau_1\otimes\sigma_3 $, and ${\mathcal R}_4=\exp\left(i\frac{\pi}{4}\tau_0\otimes\sigma_3\right)$. For a scalar (pseudoscalar) $M$ under an operation $X$ it holds $XMX^\dagger=X$ ($XMX^\dagger=-X$). Scalar and pseudoscalar operators are denoted by $\checkmark$ and $\times$ in the table. Here $\checkmark (\times)$ also indicates whether an operator is even (odd) under a specific symmetry. }
    \label{table:symm-hamiltonian}
 \end{table}
%%%%%%%%%%%%%%%%%%%%%%%%%%%%%%%%%%%%%%%%%%%%%%%%%%%%%%%%%%%%%%%%%%%%%%%%%%%%%%%%%%%%%%%%%%%%%%%%%%%%%%%%%%%%%%%%%%%%
%%%%%%%%%%%%%%%%%%%%%%%%%%%%%%%%%%%%%%%%%%%%%%%%%%%%%%%%%%%%%%%%%%%%%%%%%%%%%%%%%%%%%%%%%%%%%%%%%%%%%%%%%%%%%%%%%%%%
%%%%%%%%%%%%%%%%%%%%%%%%%%%%%%%%%%%%%%%%%%%%%%%%%%%%%%%%%%%%%%%%%%%%%%%%%%%%%%%%%%%%%%%%%%%%%%%%%%%%%%%%%%%%%%%%%%%%
%%%%%%%%%%%%%%%%%%%%%%%%%%%%%%%%%%%%%%%%%%%%%%%%%%%%%%%%%%%%%%%%%%%%%%%%%%%%%%%%%%%%%%%%%%%%%%%%%%%%%%%%%%%%%%%%%%%%
%%%%%%%%%%%%%%%%%%%%%%%%%%%%%%%%%%%%%%%%%%%%%%%%%%%%%%%%%%%%%%%%%%%%%%%%%%%%%%%%%%%%%%%%%%%%%%%%%%%%%%%%%%%%%%%%%%%%

\section{Unitary and anti-unitary particle-hole symmetry}~\label{secSI:particlehole}

In this Appendix we discuss the unitary and anti-unitary particle-hole symmetry of both first-order and second-order topological insulators that respectively protects zero-energy edge, surface, corner and hinge modes, and both zero and finite energy dislocation modes. For this analysis we choose  the following representation of the $\Gamma$ matrices satisfying the anticommuting Clifford algebra
\begin{equation}~\label{eq:gammac4breaking}
\Gamma_1=\tau_1\otimes\sigma_1, \Gamma_2=\tau_1\otimes\sigma_2, \Gamma_3=\tau_1\otimes\sigma_3, \Gamma_4=\tau_3\otimes\sigma_0, \Gamma_5=\tau_2\otimes\sigma_0,
\end{equation}
in the Hamiltonian for the second-order HOT insulator in $d=2$ and $d=3$, given by Eq.~(1) in the main text.  \\

Next we discuss the particle-hole or the spectral symmetry of both the first-order and second-order topological insulators in $d=2$ and $d=3$. It is defined in terms of an operator, say $\Theta$, which \emph{anticommutes} with a generic Hamitlonian $\hat{h}_{\rm gen}$, i.e., $\{ \hat{h}_{\rm gen}, \Theta \}=0$. If such operator ($\Theta$) exists, then all the eigenstates of the $\hat{h}_{\rm gen}$ at positive and negative energies $\pm E_n$, denoted by $| \pm E_n \rangle$, are related to each other according to $\Theta | \pm E_n \rangle =|\mp E_n \rangle$. This statement follows from the relation
\begin{eqnarray}
\hat{h}_{\rm gen} | \pm E_n \rangle = \pm E_n | \pm E_n \rangle
\Rightarrow \Theta \left( \hat{h}_{\rm gen} | \pm E_n \rangle \right) &=& \pm E_n \left( \Theta | \pm E_n \rangle \right)
\Rightarrow - \hat{h}_{\rm gen} \left( \Theta  | \pm E_n \rangle \right)= \pm E_n \left( \Theta | \pm E_n \rangle \right) \nonumber \\
&\Rightarrow& \hat{h}_{\rm gen} \left( \Theta  | \pm E_n \rangle \right)= \mp E_n \left( \Theta | \pm E_n \rangle \right),
\end{eqnarray}
and hence $\Theta | \pm E_n \rangle =|\mp E_n \rangle$. Otherwise, the particle-hole symmetry generator $\Theta$ can be either unitary or antiunitary~\cite{roy-prr2019}. The above definition of the particle-hole or spectral symmetry then shows that if there exist any state at precise zero energy, which can only be achieved in the true thermodynamic limit, then it must be an eigenstate of $\Theta$ with eigenvalue $+1$ or $-1$. However, in any finite system, all the states are at finite energies, and the states which reside at zero energy in the thermodynamic limit are also placed at finite but extremely small (typically $\sim 10^{-6}-10^{-8}$) energies. Next we show that such particle-hole or spectral symmetry operator always exists for the universal models of both first-order and second-order topological insulators in $d=2$ and $d=3$ [see Eq.~(1)], and thereby protects both zero energy topological boundary modes (such as the edge, surface, corner and hinge modes), as well as the dislocation modes (both zero and finite energy ones). \\

We begin with the first-order topological insulators. Notice that for such systems in both $d=2$ and $d=3$, we can always find a unitary particle-hole symmetry operator, namely $\Theta=\Gamma_5$, irrespective of the representation of the $\Gamma$ matrices. We note that with the representation of the $\Gamma$ matrices, shown in Eq.~(\ref{eq:gammac4breaking}), we also also find one antiunitary particle-hole symmetry generator, namely $\Theta=(\tau_2 \sigma_2) {\mathcal K}$, for 3D first-order topological insulator, where ${\mathcal K}$ is the complex conjugation and $\Theta^2=+1$. Such antiunitary particle-hole operator will play an important role for 3D HOT insulator. For now, one can immediately appreciate that the unitary particle-hole operator $\Gamma_5$, pins the boundary edge and surface modes at zero-energy and prevents them from mixing with the bulk states. In addition, it also forbids the gapless dislocation modes in both 2D and 3D first-order topological insulators from mixing with the bulk mode. Next, we proceed to HOT insulators.
\\

For a 2D HOT insulator, we can still find $\Theta=\Gamma_5$. It protects the four zero-energy corner states [see Fig.~1(b)]. Furthermore, given that the in-gap, but finite- or zero-energy dislocation modes are always well separated from the bulk states, see Fig.~\ref{FigSM:soectralflow2D}, and appear in pairs, such unitary particle-hole operator also prevents the finite-energy (for any $\theta \neq \pi/2$) and the zero-energy (for $\theta=\pi/2$) dislocation modes from mixing with the bulk states. Thus neither the zero nor the finite energy dislocation modes can be removed from the system, and they are protected by the particle-hole symmetry. Finally, for a 3D HOT insulator, we cannot find any unitary particle-hole operator, since the corresponding Hamiltonian operator exhausts all five mutually anticommuting four-component $\Gamma$ matrices. Nevertheless, the particle-hole symmetry is then generated by an antiunitary operator $\Theta=(\tau_2 \sigma_2) {\mathcal K}$. Then the above discussion for the 2D HOT insulator immediately generalizes to 3D HOT insulator, and one can immediately conclude that (1) the four zero-energy hinge modes, and (2) both finite- and zero-energy dislocation modes are protected by the antiuntary particle-hole symmetry. Finally, we note that the existence of such an antiunitary particle-hole symmetry generator is also independent of the explicit representation of the $\Gamma$ matrices~\cite{roy-prr2019}. In Appendix~\ref{app:symmetry-protection}, we show that composite symmetries of both 2D and 3D HOT insulators forbid addition of any new term to their corresponding universal Hamiltonian, shown in Eqs.~(\ref{eq:lattice-hamiltonian-2DHOT}) and (\ref{eq:3DHOTHamltonian}), respectively. Therefore, both finite and zero energy dislocation modes are symmetry protected.

%%%%%%%%%%%%%%%%%%%%%%%%%%%%%%%%%%%%%%%%%%%%%%%%%%%%%%%%%%%%%%%%%%%%%%%%%%%%%%%%%%%%%%%%%%%%%%%%%%%%%%%%
%%%%%%%%%%%%%%%%%%%%%%%%%%%%%%%%%%%%%%%%%%%%%%%%%%%%%%%%%%%%%%%%%%%%%%%%%%%%%%%%%%%%%%%%%%%%%%%%%%%%%%%%
%%%%%%%%%%%%%%%%%%%%%%%%%%%%%%%%%%%%%%%%%%%%%%%%%%%%%%%%%%%%%%%%%%%%%%%%%%%%%%%%%%%%%%%%%%%%%%%%%%%%%%%%
%%%%%%%%%%%%%%%%%%%%%%%%%%%%%%%%%%%%%%%%%%%%%%%%%%%%%%%%%%%%%%%%%%%%%%%%%%%%%%%%%%%%%%%%%%%%%%%%%%%%%%%%
%%%%%%%%%%%%%%%%%%%%%%%%%%%%%%%%%%%%%%%%%%%%%%%%%%%%%%%%%%%%%%%%%%%%%%%%%%%%%%%%%%%%%%%%%%%%%%%%%%%%%%%%
\begin{table}
    \begin{tabular}{|c|c|c|c|c|c|c|c|}
    \hline
    \multicolumn{2}{|c|}{Operator}& \multicolumn{6}{|c|}{Symmetry} \\ \hline
    Compact form & Explicit form & ${\mathcal T}$ & ${\mathcal P}$ & $C_4$& ${\mathcal P}{\mathcal T}$ & $C_4{\mathcal P}$ & $C_4{\mathcal T}$ \\
    \hline
    $\Gamma_0$ & $\tau_0\sigma_0$ & $\checkmark$ & $\checkmark$  & $\checkmark$  & $\checkmark$  & $\checkmark$  & $\checkmark$ \\ \hline
    $\Gamma_3$ & $\tau_1\sigma_3$ & $\checkmark$  & $\checkmark$  & $\checkmark$  & $\checkmark$  & $\checkmark$  & $\checkmark$ \\ \hline
    $\Gamma_4$ & $\tau_3\sigma_0$ & $\times$ & $\times$ & $\checkmark$ & $\checkmark$ & $\times$ & $\times$\\ \hline
    $\Gamma_5$ & $\tau_2\sigma_0$ & $\times$ & $\times$ & $\checkmark$ & $\checkmark$ & $\times$ & $\times$\\ \hline
    $\Gamma_{12}$ & $-\tau_0\sigma_3$ & $\times$ & $\checkmark$ & $\checkmark$ & $\times$ & $\checkmark$ & $\times$\\ \hline
    $\Gamma_{34}$ & $\tau_2\sigma_3$ & \checkmark & $\times$ & $\checkmark$  & $\times$ & $\times$ & $\checkmark$\\ \hline
    $\Gamma_{35}$ & $-\tau_3\sigma_3$ & \checkmark & $\times$ & $\checkmark$  & $\times$ & $\times$ & $\checkmark$\\ \hline
    $\Gamma_{45}$ & $\tau_1\sigma_0$ & $\times$ & $\checkmark$ & $\checkmark$ & $\times$ & $\checkmark$ & $\times$\\ \hline
    $Y^1=(\Gamma_1,\Gamma_2)$ & $(\tau_1\sigma_1,\tau_1\sigma_2)$ & $\times$ & $\times$ & $\checkmark$ & $\checkmark$ & $\times$ & $\times$\\ \hline
    $Y^2=(\Gamma_{23},\Gamma_{13})$ & $(\tau_0\sigma_1,\tau_0\sigma_2)$ & $\checkmark$ & $\times$ & $\checkmark$ & $\times$ & $\times$ & $\checkmark$\\ \hline
    $Y^3=(\Gamma_{14},\Gamma_{24})$ & $(\tau_2\sigma_1,\tau_2\sigma_2)$ & $\times$ & $\checkmark$ & $\checkmark$ & $\times$ & $\checkmark$ & $\times$\\ \hline
    $Y^4=(\Gamma_{15},\Gamma_{25})$ & $(-\tau_3\sigma_1,-\tau_3\sigma_2)$ & $\times$ & $\checkmark$ & $\checkmark$ & $\times$ & $\checkmark$ & $\times$\\ \hline
    \end{tabular}
    \caption{Symmetry transformation of the identity matrix ($\Gamma_0$), five $\Gamma$ matrices in Eq.~\eqref{eq:gammac4breaking} and ten commutators $\Gamma_{jk}=[\Gamma_j,\Gamma_k]/(2i)$ under time-reversal (${\mathcal T}$),  parity (${\mathcal P}$) and $C_4$ rotational (${\mathcal R}_4$) symmetries, which  are, respectively, represented by the operators ${\mathcal T}=\tau_2\otimes\sigma_1 {\mathcal K}$, ${\mathcal P}=\tau_1\otimes\sigma_3 $, and ${\mathcal R}_4=\exp\left(i\frac{\pi}{4}\tau_0\otimes\sigma_3\right)$. Scalar and vector operators are denoted by $\checkmark$, while $\times$ denotes pseudoscalar and pseudovector operators. Here $\checkmark (\times)$ also indicates whether an operator is even (odd) under a specific symmetry.}
    \label{table:Gamma-matrices}
 \end{table}
%%%%%%%%%%%%%%%%%%%%%%%%%%%%%%%%%%%%%%%%%%%%%%%%%%%%%%%%%%%%%%%%%%%%%%%%%%%%%%%%%%%%%%%%%%%%%%%%%%%%%%%%
%%%%%%%%%%%%%%%%%%%%%%%%%%%%%%%%%%%%%%%%%%%%%%%%%%%%%%%%%%%%%%%%%%%%%%%%%%%%%%%%%%%%%%%%%%%%%%%%%%%%%%%%
%%%%%%%%%%%%%%%%%%%%%%%%%%%%%%%%%%%%%%%%%%%%%%%%%%%%%%%%%%%%%%%%%%%%%%%%%%%%%%%%%%%%%%%%%%%%%%%%%%%%%%%%
%%%%%%%%%%%%%%%%%%%%%%%%%%%%%%%%%%%%%%%%%%%%%%%%%%%%%%%%%%%%%%%%%%%%%%%%%%%%%%%%%%%%%%%%%%%%%%%%%%%%%%%%
%%%%%%%%%%%%%%%%%%%%%%%%%%%%%%%%%%%%%%%%%%%%%%%%%%%%%%%%%%%%%%%%%%%%%%%%%%%%%%%%%%%%%%%%%%%%%%%%%%%%%%%%

\section{Details of the numerical analysis and additional results}~\label{app:D}

This Appendix is devoted to highlight the parameter values used in Figs.~1 and 2 of the main text, and discuss additional numerical results for three-dimensional second-order topological insulators, hosting one-dimensional hinge modes along the $z$ direction, in the presence of edge and screw dislocation-antidislocation pair, shown in Fig.~\ref{FigSM:3Ddislocation}. \\

For two-dimensional topological insulators, we always choose $t=1$,  $B=1/2$, and $m=3$ [see Eq.~\eqref{eq:lattice-hamiltonian-2DHOT}], so that the system is in the $M$ phase. To realize a second-order topological insulator, we set $\Delta=0.20$ in Figs.~1(b) and 1(c). For  Figs.~2(a)-(c), we consider a dislocation-antidislocation pair  in a periodic system with linear dimension $L=28$ in each direction. The locations of the dislocations in these figures are identical to the ones in Fig.~1(c). \\

For three-dimensional topological insulators, we always set $t=B=1$ and $m=10$ [see Eq.~\eqref{eq:3DHOTHamltonian}], so that the system is in the $R$ phase. To realize a second-order topological insulator, we set $\Delta=0.40$ in Fig.~3.
\\

We now discuss additional numerical results for 3D second-order topological insulators in the presence of dislocations, displayed in Fig.~\ref{FigSM:3Ddislocation}. In Figs.~\ref{FigSM:3Ddislocation}(a)-(c) we show the scaling of the spectral gap ($\delta E$) among four states localized in the dislocation core in the presence of an edge dislocation-antidislocation pair with Burgers vectors ${\bf b}=\pm a {\bf e}_x$. It shows that only when the Burgers vector is parallel to one of the $C_4$ symmetry breaking axis ($\theta=\pi/2$) these modes become gapless [Fig.~\ref{FigSM:3Ddislocation}(a)]. Otherwise, the spectral gap scales linearly with $\cos \theta$ and $\Delta$, as shown in Figs.~\ref{FigSM:3Ddislocation}(b) and (c), respectively, when $\Delta$ is small, in agreement with our scaling argument presented in Appendix~\ref{SMsec:3Ddislocation}.   \\

Figs.~\ref{FigSM:3Ddislocation}(d) and (e) show the localization (through LDoS) of the dislocation modes around the core of edge and screw dislocation-antidislocation pair, respectively, in the $xy$ plane for a specific $z=6$. These two figures are projections of Figs.~3(b) and (d) of the main text, respectively, for a specific $z$. \\

Fig.~\ref{FigSM:3Ddislocation}(f) shows the scaling of the spectral gap ($\delta E$) between the states localized in the dislocation core in the presence of a screw dislocation-antidislocation pair for two specific orientations of the corresponding Burgers vectors, namely, when ${\bf b}=\pm a {\bf e}_z$ (red) and ${\bf b}=\pm a {\bf e}_y$ (blue). For ${\bf b}=\pm a {\bf e}_z$ the screw dislocation pair pierces the $xy$ surfaces that host gapless modes, while for ${\bf b}=\pm a {\bf e}_y$ it pierces the $xz$ planes which, on the other hand, accommodate gapped modes. Concomitantly, for these two orientations of the screw dislocation-antidislocation pair the modes localized in its core are respectively gapless and gapped, in agreement with our analytical arguments from Appendix~\ref{SMsec:3Ddislocation}.\\

Finally, we show the real space version of the tight-binding model, reported in Eq.~(1) of the main text in the momentum space, in the presence of a single edge dislocation in open system on a 2D square lattice for $\theta=0$. By performing the Fourier transform, we arrive at
\allowdisplaybreaks[4]
\begin{align}
\hat{h}_{\rm real} &= \bigg( \bigg[ \left\{
\left( \sum^{\ell_x-2}_{n_x=1} \sum^{\ell_y-1}_{n_y=1}
+ \sum^{L_x-1}_{n_x=\ell_x+1} \sum^{\ell_y-1}_{n_y=1}
+ \sum^{L_x-1}_{n_x=1} \sum^{L_y}_{n_y=\ell_y} \right) c^\dagger_{n_x,n_y} c_{n_x+1,n_y}
+ \sum^{\ell_y}_{n_y=1} c^\dagger_{\ell_x-1,n_y} c_{\ell_x+1,n_y}
\right\} \otimes \frac{2 B \Gamma_3 + i t \Gamma_1 + \Delta \Gamma_4}{2} \nonumber \\
&+ \left\{
\left( \sum^{\ell_x-1}_{n_x=1} \sum^{L_y-1}_{n_y=1} + \sum^{L_x}_{n_x=\ell_x+1} \sum^{L_y-1}_{n_y=1} \right) c^\dagger_{n_x,n_y} c_{n_x,n_y+1}
+ \sum^{L_y-1}_{n_y=\ell_y} c^\dagger_{\ell_x,n_y} c_{\ell_x,n_y+1}
\right\} \otimes \frac{2 B \Gamma_3 + i t \Gamma_2 - \Delta \Gamma_4}{2} \bigg] + H.c. \bigg)\nonumber \\
&+ \sum^{L_x L_y-(\ell_y-1)}_{j=1} c^\dagger_{j,j} c_{j,j} \otimes (m-4 B) \Gamma_3.
\end{align}
In this construction, the dislocation center is located at $(\ell_x,\ell_y)$ [see Fig.~1 of the main text]. Now this construction can be generalized to introduce a pair of edge dislocation-antidislocation in a 2D periodic system, as well as edge and screw dislocations in 3D.

\section{Symmetry protection of dislocation modes}~\label{app:symmetry-protection}

In this Appendix we take the representation of the $\Gamma$ matrices as given by Eq.~\eqref{eq:gammac4breaking} in the Hamiltonian for the second-order HOT insulator in $d=2$ and $d=3$, given by Eq.~(1) in the main text. Antiunitary  time-reversal (${\mathcal T}$), unitary parity (${\mathcal P}$) and $C_4$ rotational (${\mathcal R}_4$) symmetries are respectively represented by the operators ${\mathcal T}=\tau_2\otimes\sigma_1 {\mathcal K}$, ${\mathcal P}=\tau_1\otimes\sigma_3 $, and ${\mathcal R}_4=\exp\left(i\frac{\pi}{4}\tau_0\otimes\sigma_3\right)$, where ${\mathcal K}$ is the complex conjugation. Under time-reversal and parity, momentum changes sign, ${\bf k}\rightarrow-{\bf k}$, while it is a vector under the $C_4$ rotation: $(k_x,k_y)\rightarrow(-k_y,k_x)$. The first-order topological Hamiltonian ($\hat{h}_0$) and the second-order mass term ($\hat{h}_\Delta$) are both invariant \emph{only} under the composite ${\mathcal P}{\mathcal T}$, ${C}_4{\mathcal T}$, ${C}_4{\mathcal P}$ symmetries, see Table~\ref{table:symm-hamiltonian}. We show that there are no additional symmetry allowed terms that can be added to the HOT Hamiltonian besides the ones already present and the trivial one, proportional to unity matrix, which is, of course, always symmetry allowed in an insulating system, and it only causes an overall shift of energy eigenvalues. Finally, we point out that this proof pertains to any second-order topological insulator with $C_{4n}$ rotational symmetry breaking second-order Wilson-Dirac mass because in this case for any $n$ the HOT mass is \emph{even} under ${\bf k}\to -{\bf k}$. \\

As a first step, we classify all the sixteen four-dimensional Hermitian matrices, constructed from five mutually anticommuting Hermitian $\Gamma$ matrices given by Eq.~\eqref{eq:gammac4breaking}, according to the above three composite symmetries, see Table~\ref{table:Gamma-matrices}. We see that only the matrix $\Gamma_3$, which represents the first-order topological mass, is invariant under all the three composite or product symmetry operations. One can also form scalars under the $C_4$ rotation in terms of the four vectors in the form,
\begin{equation}\label{eq:Hj-scalar}
H_V^{j}=\Delta_j(\cos\chi Y^j_1+\sin\chi Y^j_2),
\end{equation}
where $j=1,2,3,4$ corresponds to four vectors $Y^j$ from the last four rows in Table~\ref{table:Gamma-matrices}, and the angle $\chi\rightarrow\chi+\pi/2$ under the $C_4$. However, as explicitly shown in Table~\ref{table:vector-Hamiltonian}, none of these Hamiltonians is invariant under \emph{both}
 parity and time-reversal, and therefore they violate either $C_4{\mathcal T}$, $C_4{\mathcal P}$, or ${\mathcal P}{\mathcal T}$. Notice that the ${\mathcal P}{\mathcal T}$ symmetry assures the double degeneracy of the electronic bands in the model, since it is antiunitary and squares to $-1$~\cite{roy-prr2019}. Therefore, there are no new symmetry allowed terms in the 2D HOT Hamiltonian given by Eq.~(1) in the main text besides the ones already present. This in turn implies that both both finite and zero energy dislocation modes in a 2D second-order topological insulator are symmetry protected by the same combinations of the spatial and non-spatial symmetries as the bulk Hamiltonian. \\

In three spatial dimensions, we choose the first order Hamiltonian so that the matrix $\Gamma_{5}$ multiplies the factor $t \sin (k_z a)$ (see Table~\ref{table:Gamma-matrices}), while we keep the rest of the terms identical as in the Hamiltonian for the 2D second-order topological insulator [a slightly different notation than the one in Eq.~(1)]. Therefore, requiring the same symmetries, the above results imply that the same composite symmetries protect the dislocation modes in 3D as in two spatial dimensions.

In Fig.~\ref{FigSM:soectralflow2D}, we show that in the presence of dislocations the energy spectra possess two gaps. One is set by the first-order topological mass (among the black colored bulk states) and the other one is determined by the HOT mass $\Delta({\bf k}, \theta)$ among the dislocation modes (red states). The latter one in addition also depends on the parameter $\theta$, measuring the relative orientation between the HOT mass domain walls and the Burgers vector. Only when $\theta=\pi/2$, for which the HOT mass domain wall is parallel to the Burgens vector, the dislocation modes become gapless. However, as we argued above, these two distinct energy scales are symmetry protected and we cannot add any symmetry allowed perturbation that can mix these two scales. The same conclusion holds analogously in 3D. Thus both finite and zero energy dislocation modes are symmetry protected.

%%%%%%%%%%%%%%%%%%%%%%%%%%%%%%%%%%%%%%%%%%%%%%%%%%%%%%%%%%%%%%%%%%%%%%%%%%%%%%%%%%%%%%%%%%%%%%%%%%%%%%%%
%%%%%%%%%%%%%%%%%%%%%%%%%%%%%%%%%%%%%%%%%%%%%%%%%%%%%%%%%%%%%%%%%%%%%%%%%%%%%%%%%%%%%%%%%%%%%%%%%%%%%%%%%
\section{Symmetry protection for $C_{4n+2}$ symmetry breaking second-order topological insulator}
\label{app:F}

In this Appendix, we discuss the symmetry protection of the dislocation modes when the rotational symmetry is of the form $C_{4n+2}$, where $n$ is an integer, as for instance is the case for $C_6$ rotations. Specifically for $C_6$ symmetry breaking HOT insulators the second-order Wilson-Dirac mass takes the form 
\begin{equation}
\hat{h}_\Delta=\Delta \Gamma_{d+2} \left[ \cos \theta \; \sin(k_x a) \left\{ 3 \cos(k_y a)-\cos(k_x a) + 2 \right\}
+ \sin \theta \; \sin(k_y a) \left\{ 3 \cos(k_x a)-\cos(k_y a) + 2 \right\} \right],
\end{equation}
in the $M$ phase [compare with Eq.~(\ref{eq:lattice-hamiltonian-HOT})]~\cite{roy-goswami-juricic}. In this case, the second-order Wilson-Dirac mass is \emph{odd} under ${\bf k}\to -{\bf k}$, and the protection of the dislocation modes can be realized only through the parity (${\mathcal P}$) and the time-reversal (${\mathcal T}$) symmetries (no rotational or a composite symmetry is required now). Therefore, the inspection of the Table~\ref{table:Gamma-matrices}, shows that the only $\Gamma$ matrices preserving both parity and time-reversal symmetries are $\Gamma_0$ and $\Gamma_3$. Then our argument from Appendix~\ref{app:symmetry-protection} can be extended straightforwardly to show that dislocation modes (at both finite and zero energies) cannot be mixed with the other bulk states and hence they are symmetry protected. Furthermore, this argument straightforwardly generalizes to all $2D$ second-order topological insulators for which the second order Dirac-Wilson mass is odd under the transformation ${\bf k}\to-{\bf k}$. Finally, due to the form of the Hamiltonian for 3D second-order topological insulators, this argument straightforwardly applies also to this case.

%%%%%%%%%%%%%%%%%%%%%%%%%%%%%%%%%%%%%%%%%%%%%%%%%%%%%%%%%%%%%%%%%%%%%%%%%%%%%%%%%%%%%%%%%%%%%%%%%%%%%%%%%%%%%%%%%%%%%%%%%%%%%%%%%%
%%%%%%%%%%%%%%%%%%%%%%%%%%%%%%%%%%%%%%%%%%%%%%%%%%%%%%%%%%%%%%%%%%%%%%%%%%%%%%%%%%%%%%%%%%%%%%%%%%%%%%%%%%%%%%%%%%%%%%%%%%%%%%%%%%
\begin{table}
    \begin{tabular}{|c|c|c|c|c|c|c|}
    \hline
    \multirow{2}{*}{Operator}& \multicolumn{6}{|c|}{Symmetry}\\
    \cline{2-7}
     & ${\mathcal T}$ & ${\mathcal P}$ & $C_4$& ${\mathcal P}{\mathcal T}$ & $C_4{\mathcal P}$ & $C_4{\mathcal T}$ \\
    \hline
    $H_V^{1}$  & $\times$ & $\times$ & $\checkmark$ & $\checkmark$ & $\times$ & $\times$\\
    \hline
    $H_V^{2}$  & $\checkmark$ & $\times$ & $\checkmark$ & $\times$ & $\times$ & $\checkmark$\\
    \hline
    $H_V^{3}$  & $\times$ & $\checkmark$ & $\checkmark$ & $\times$ & $\checkmark$ & $\times$\\
    \hline
    $H_V^{4}$  & $\times$ & $\checkmark$ & $\checkmark$ & $\times$ & $\checkmark$ & $\times$\\
    \hline
    \end{tabular}
    \caption{Symmetry transformations of the Hamiltonians defined in Eq.~\eqref{eq:Hj-scalar}. The notation is the same as in the Tables~\ref{table:symm-hamiltonian} and \ref{table:Gamma-matrices}. Notice that none of the Hamiltonians is invariant under all the three composite symmetries. Here $\checkmark (\times)$ indicates whether an operator is even (odd) under a specific symmetry.}
    \label{table:vector-Hamiltonian}
 \end{table}
%%%%%%%%%%%%%%%%%%%%%%%%%%%%%%%%%%%%%%%%%%%%%%%%%%%%%%%%%%%%%%%%%%%%%%%%%%%%%%%%%%%%%%%%%%%%%%%%%%%%%%%%%%%%%%%%%%%%%%%%%%%%%%%%%%
%%%%%%%%%%%%%%%%%%%%%%%%%%%%%%%%%%%%%%%%%%%%%%%%%%%%%%%%%%%%%%%%%%%%%%%%%%%%%%%%%%%%%%%%%%%%%%%%%%%%%%%%%%%%%%%%%%%%%%%%%%%%%%%%%%

\twocolumngrid

\end{document}